\documentclass[prd,preprint,tightenlines,floatfix,showpacs,preprintnumbers,nofootinbib,eqsecnum]{revtex4-1}

\pdfoutput=1

\usepackage[T1]{fontenc} 
\usepackage[utf8]{inputenc}

\usepackage{url}
\usepackage{epsfig}
\usepackage{amsmath}
\usepackage{amssymb}
\usepackage{amsfonts}
\usepackage{bbm}
\usepackage[footnotesize]{caption}
\usepackage{graphicx}
\usepackage{mathrsfs}
\usepackage{color}
\usepackage[dvipsnames]{xcolor}
\usepackage{comment}
\usepackage{mathtools}
\usepackage{braket}

\usepackage[caption=false,justification=justified]{subfig}  
\usepackage{todonotes}

\usepackage{hyperref}
\usepackage{multirow}
\usepackage{relsize}
\usepackage{fullpage}
\usepackage{makecell}
\usepackage{enumitem,xcolor}

\usepackage{xspace}
\usepackage{bm}         
\usepackage{array}
\usepackage{booktabs}   
\usepackage{mleftright}

\usepackage{bbold}
\usepackage{amscd}
\usepackage{soul}
\usepackage{tikz-cd}
\usepackage{empheq}
\usepackage[most]{tcolorbox}
\usepackage{slashed}
\usepackage{dsfont}
\usepackage[makeroom]{cancel}

\usepackage[capitalise]{cleveref}

\crefname{section}{Sec.}{Secs.}
\crefname{table}{Tab.}{Tabs.}
\crefname{figure}{Fig.}{Figs.}
\crefname{equation}{Eq.}{Eqs.}
\crefname{appendix}{Appendix\ }{Appendix\ }

\newcommand{\U}[1]{\mathrm{U}(1)_{\mathrm{#1}}}			
\newcommand{\SU}[2]{\mathrm{SU}(#1)_{\mathrm{#2}}}		
\newcommand{\E}[1]{\mathrm{E}_{#1}}	

\newcommand{\blue}[0]{\color{blue}}

\newcommand{\red}[0]{\color{red}}

\begin{document}

\title{Gauge couplings evolution from the Standard Model, through Pati-Salam theory, into $\E{8}$ unification of families and forces}

\author{Francisco~J.~de~Anda}
\email{fran@tepaits.mx}
\affiliation{Tepatitl{\'a}n's Institute for Theoretical Studies, C.P. 47600, Jalisco, M{\'e}xico}

\author{Alfredo Aranda}
\email{fefo@ucol.mx}
\affiliation{Facultad de Ciencias-CUICBAS, Universidad de Colima, C.P.28045, Colima, M\'exico 01000, M\'exico}
\affiliation{Dual CP Institute of High Energy Physics, C.P. 28045, Colima, M\'exico}

\author{Ant{\'o}nio~P.~Morais}
\email{aapmorais@ua.pt}
\affiliation{Departamento de F\'isica, Universidade de Aveiro and CIDMA, Campus de Santiago, 
3810-183 Aveiro, Portugal}
\affiliation{Department of Astronomy and Theoretical Physics, Lund University, 221 00 Lund, Sweden}

\author{Roman~Pasechnik}
\email{Roman.Pasechnik@thep.lu.se}
\affiliation{Department of Astronomy and Theoretical Physics, Lund University, 221 00 Lund, Sweden\vspace{1cm}}

\begin{abstract}
\vspace{0.5cm}
We explore the potential of ultimate unification of the Standard Model matter and gauge sectors into a single 
$E_8$ superfield in ten dimensions via an intermediate Pati-Salam gauge theory. Through a consistent realisation of a $\mathbb{T}^6/(\mathbb{Z}_6\times \mathbb{Z}_2)$ orbifolding procedure accompanied by the Wilson line breaking mechanism and Renormalisation Group evolution of gauge couplings, we have established several benchmark 
scenarios for New Physics that are worth further phenomenological exploration.
\end{abstract}

\maketitle


\section{Introduction}
\label{sec:intro}

Grand Unified Theories (GUTs) aim to unify the three independent gauge interactions of the Standard Model (SM) into a single one. Among the simplest ways to achieve this is via an $SU(5)$ gauge symmetry \cite{Georgi:1974sy}. One can also extend the gauge symmetry to unify all the SM fermions of a single family into one $SO(10)$ representation \cite{Fritzsch:1974nn}. Enlarging the gauge group even further into $E_6$, one could unify the SM Higgs sector and a full family of fermions into a single representation \cite{King:2005my} by means of the simple ${\cal N}=1$ supersymmetry (SUSY). 

Such a step-by-step enlarging of the gauge symmetry group can be studied through its Dynkin diagram and is known to follow the exceptional chain \cite{Buchmuller:1985rc,Koca:1982zi},
\begin{equation}
 SU(3)_C\times SU(2)_L\times U(1)_Y \subset SU(5) \subset SO(10) \subset E_6 
 \subset E_7 \subset E_8 \,. \label{Eq:chain}
\end{equation}
Note, the group $E_8$ is the largest group of the chain and is specially relevant as its adjoint representation $(\textbf{248})$ is the same as the fundamental one \cite{Slansky:1981yr}. This suggests that the SM gauge fields can be in principle unified with the SM fermions provided that a maximal $\mathcal{N}=4$ SUSY is realised. Furthermore, it also provides an $SU(3)$ flavor (or family) symmetry as a coset of $E_8$ to $E_6$ reduction.

There is a plethora of models in the literature that aim to build GUTs including flavor symmetries \cite{King:2001uz,King:2017guk,Hagedorn:2010th,Antusch:2014poa,Bjorkeroth:2015ora,Bjorkeroth:2015uou,Bjorkeroth:2017ybg,deAnda:2017yeb,CarcamoHernandez:2020owa,Morais:2020ypd,Morais:2020odg,Camargo-Molina:2016yqm,Camargo-Molina:2017kxd,Camargo-Molina:2016yqm,Camargo-Molina:2016bwm} and extra dimensions (EDs), \cite{Altarelli:2008bg,Burrows:2009pi,Burrows:2010wz,deAnda:2018oik,Altarelli:2006kg,Adulpravitchai:2010na,Adulpravitchai:2009id,Asaka:2001eh,deAnda:2019jxw,deAnda:2018ecu,deAnda:2018yfp}). In order to achieve viability, most require a number of independent groups and quite a large number of fields. As alluded above, the group $E_8$ seems to be a good bet for a complete unification of SM vectors, fermions and scalars and has indeed been widely studied both in the context of string theory \cite{Ibanez:1987pj,Parr:2020oar} and within the framework of quantum field theory (QFT) \cite{Adler:2002yg,Adler:2004uj,Garibaldi:2016zgm,Thomas:1985be,Konshtein:1980km,Baaklini:1980fv,Baaklini:1980uq,Barr:1987pu,Bars:1980mb,Koca:1981xd,Mahapatra:1988gc,Ong:1984ej,Camargo-Molina:2016yqm,Olive:1982ai}.

The first challenge posed by $E_8$ is that it is a real group (as is extended SUSY) while the SM requires chiral representations. A way forward is to assume the existence of extra dimensions which are orbifolded in such a way that, after their compactification, the massless chiral representations containing the SM fermion sectors remain. Furthermore, extended SUSY can also be generated from the EDs \cite{ArkaniHamed:2001tb,Brink:1976bc}. A big challenge is to avoid the presence of too many massless states in the low-energy limit of the theory typically originating by an orbifolding procedure of $E_8$. A consistent resolution may be found by a combination of orbifolding and the Wilson line symmetry-breaking mechanism, also associated to the specific orbifold structure, providing a reduction of the symmetry group and generation of large masses for many of the unobserved states.

One of the main requirements for a consistent GUT is gauge couplings unification, i.e., when the SM gauge couplings are evolved from their measured values at the electroweak (EW) scale up to the high-energy scales and they match into a single coupling of a unified gauge group. Recently~\cite{Aranda:2020noz,Aranda:2020zms}, some of us presented a framework where the $E_8$ gauge symmetry with $\mathcal{N}=1$ SUSY is considered in 10d corresponding to an extended $\mathcal{N}=4$ SUSY in 4d, where the EDs are orbifolded so that after the compactification stage only simple SUSY and Pati-Salam symmetry remain \cite{Pati:1973uk}.

In this paper, we study a particular realisation of the GUT with $E_8$ gauge symmetry in 10d, where the full SM (Higgs, gauge, fermion) field content is unified into a single $E_8$ gauge superfield. Through the $\mathbb{T}^6/(\mathbb{Z}_6\times \mathbb{Z}_2)$ orbifold, the $E_8$ symmetry is broken down into $SU(4)_{PS}\times SU(2)_L\times SU(2)_R\times U(1)_X\times U(1)_F\times SU(2)_F$. In order to consistently derive the 4d Pati-Salam theory from $E_8$ in 10d, the ED compactification procedure invokes the presence of several extra fields for the model to be anomaly free at every step\footnote{We are indebted to Stephen F. King for important discussions on this topic at early stages of this work.}. Thus, we assume the existence of additional chiral superfields, consistent with the $E_8$ symmetry, so that the anomalies are manifestly canceled at the Pati-Salam level. The symmetry is further broken down to the SM one through Wilson lines.

The SM gauge and matter sectors originate from a single gauge superfield and remain massless up to the EW symmetry breaking scale. The additional chiral fields have arbitrary masses and thus may not be present at low energies. In this work, we study different benchmark scenarios of New Physics where some of the fields survive below the compactification scale and provide important contributions to the renormalization group (RG) evolution of the gauge couplings necessary to achieve the exact gauge couplings unification below the Planck scale.

The layout of the paper is as follows: In Sec.~\ref{sec:orb} we show the basics of the orbifolding mechanism employed in this work. In Sec.~\ref{sec:t6e8} we demonstrate how the specific $\mathbb{T}_6/(\mathbb{Z}_6\times \mathbb{Z}_2)$ orbifold breaks the $E_8$ gauge symmetry and SUSY in 10d. In Sec.~\ref{sec:anom} we introduce the extra superfields that are needed to cancel anomalies at the Pati-Salam level. In Sec.~\ref{sec:RGE} we present the benchmark scenarios relevant for further explorations of New Physics phenomenology where the exact gauge couplings unification is achieved. Finally, in Sec.~\ref{sec:con} the basic conclusions are summarised.

\section{Orbifolding}
\label{sec:orb}

Let us start by considering a theory in 10d spacetime with $\mathcal{N}=1$ SUSY. Six extra spatial dimensions are assumed to be orbifolded  as $\mathbb{T}^6/F$, where $F$ is a discrete subgroup of the extra dimensional Poincar\`e symmetry $O(6)\ltimes T^6$, such that $O(6)\simeq SO(6)\times \mathbb{Z}_2\simeq SU(4)\times \mathbb{Z}_2$ is the rotation group, while $T^6$ is the translation group. The translation group is modded by the lattice vectors $\Gamma=\mathbb{Z}^6$ compactifying it as $\mathbb{R}^6\to \mathbb{T}^6\simeq\mathbb{R}^6/\Gamma$ into a 6d torus. The group $F$ must leave the lattice invariant, i.e. $F\Gamma=\Gamma$. In order to preserve simple ${\cal N}=1$ SUSY after orbifold compactification, one should leave an invariant $U(1)$ subgroup of the rotations, therefore $F\subset SU(3)$~\cite{Dixon:1985jw,Dixon:1986jc}. 

A simple orbifolding that preserves $\mathcal{N}=1$ SUSY reads 
\begin{equation}
F\simeq \mathbb{Z}_N \subset SU(3) \,,
\end{equation}
with a positive integer $N$, where a generic $\mathbb{Z}_N$ orbifolding 
can be defined by identifying
\begin{eqnarray} \nonumber
&& (x,z_1,z_2,z_3)\sim (x,\omega_1 z_1,\omega_2 z_2,\omega_3 z_3) \,, \qquad \omega_i \equiv e^{2i \pi n_i/N} \,, \\
&& {\cal V}(x,z_1,z_2,z_3) = R(\omega_1,\omega_2,\omega_3) {\cal V}(x,\omega_1 z_1,\omega_2 z_2,\omega_3 z_3) \, .
\label{ni}
\end{eqnarray}
Here, $R$ corresponds to a representation of the $\mathbb{Z}_N$ rotation 
acting on the $10$d vector superfield ${\cal V}$. Such a transformation would 
belong to $SU(3)$ and preserve SUSY as long as
\begin{equation}
n_1+n_2+n_3=0\mod 2N \,,
\label{eq:ni}
\end{equation}
as fermions rotate twice as slow \cite{GrootNibbelink:2017luf}. 

Considering only an Abelian orbifolding that preserves $\mathcal{N}=1 $ SUSY, the most general
one reads as $F\simeq \mathbb{Z}_N \times  \mathbb{Z}_M\subset SU(3)$,
where $N,M$ are positive integers \cite{deAnda:2019anb,Fischer:2012qj,Fischer:2013qza}. 
If the boundary condition that breaks the gauge group is imposed, the orbifolding must be accompanied by an $E_8\supset U(1)_f\supset \mathbb{Z}_N$ transformation. As a result, the decomposed $10$d superfield is transformed as follows
\begin{equation}\begin{split}
V(x,z_1,z_2,z_3)&=e^{2i \pi q^f_a/N} V(x,\omega_1 z_1,\omega_2 z_2,\omega_3 z_3),\\
\phi^i(x,z_1,z_2,z_3) &=\omega_i e^{2i \pi q^f_a/N} \phi^i(x,\omega_1 z_1,\omega_2 z_2,\omega_3 z_3),
\end{split}
\label{phi^i}
\end{equation}
where $q_a^f$ is the $U(1)_f$ charge of the corresponding representation. Here, $V$ lies in the adjoint representation of the unbroken gauge group, while the light chiral superfields $\phi_i$ belong to the corresponding fundamental representation with charge $q^f_a=-n_i \mod N$. The desired light fields are
then specified by an appropriate choice of $n_i$. This fully determines the orbifolding procedure.

The underlined gauge symmetry can be broken and its rank reduced
by adding a gauge transformation to the EDs translations through the so-called Wilson line mechanism. The latter generates a mass splitting similar to the one a Vacuum Expectation Value (VEVs) would generate. It is therefore usual to parametrize this symmetry breaking by an effective VEV. It should be remembered that it is not actually a VEV, as it does not come from the minimization of a potential, but from the ED profiles of the fields, determined by boundary conditions.
In consistency with the orbifold boundary conditions, the effective VEVs should obey the rotation-translation commutation relations coming from the Poincar\`e algebra and hence emerge in chiral supermultiplets that have a zero mode. An effective potential for the fields is obtained by integrating out all the other fields (for more details, see e.g.~Refs.~\cite{Hosotani:1983xw,Hosotani:1983vn,Hosotani:2004wv,Hosotani:2004ka,Haba:2004qf,Haba:2002py}).

\section{The $\mathbb{T}^6/(\mathbb{Z}_6\times \mathbb{Z}_2)$ orbifold with $E_8$}
\label{sec:t6e8}

Now, we consider an $E_8$ gauge theory in 10d spacetime. The $E_8$ gauge symmetry has rank $8$ and the orbifolding must preserve the rank-4 SM gauge symmetry, $SU(3)_C\times SU(2)_L\times U(1)_Y$. In this work, we employ the following decomposition~\cite{Slansky:1981yr}
\begin{equation}
\begin{split}
E_8&\supset  SO(10)\times U(1)_{X'}\times SU(3)_F\\
&\quad\supset SO(10)\times U(1)_{X'}\times U(1)_{F}\times SU(2)_F \\
&\quad\supset G_{PS}\equiv SU(4)_{PS}\times SU(2)_L\times SU(2)_R\times U(1)_{X'}\times U(1)_F\times SU(2)_F \,,
\label{eq:decomp}
\end{split}
\end{equation}
and consider the $T^6/(\mathbb{Z}_6\times \mathbb{Z}_2)$ orbifold whose compactification triggers the breaking
$E_8\to G_{PS}$ i.e. featuring a Pati-Salam SUSY theory and a flavor $U(1)_{F}\times SU(2)_F$ symmetry in 4d. Subsequent reduction of this symmetry occurs in the following steps,
\begin{equation}
\begin{aligned}
& SU(4)_{PS} \to SU(3)_C \times U(1)_{B-L} \,, \qquad SU(2)_R \to U(1)_{T_3^R} \,, \\
& SU(4)_{PS} \times SU(2)_R \to SU(3)_C \times U(1)_{Y} \times U(1)_{X} \,,
\end{aligned}
\end{equation}
with
\begin{equation}
q_Y=6q_{T_3^R} + 3q_{B-L} \,, \qquad q_X=4Q_{T_3^R} - 3q_{B-L} \,,
\end{equation}
where one recovers the color $SU(3)_C$ and hypercharge $U(1)_{Y}$ groups of the SM.

The orbifold boundary conditions which provide such a breaking pattern read
\begin{equation}
\mathbb{Z}_6:\ \phi \to e^{2i\pi q_{X'}/6}\phi \,,\ \ \  
\mathbb{Z}_2:\ \phi \to e^{2i\pi (q_{Y}+q_{T_8^F})/2}\phi \,,
\label{eq:orbcondi}
\end{equation}
 with the orbifold transformations
\begin{equation}
\begin{split}
(x,z_1,z_2,z_3) &\sim (x,\alpha^2 z_1,\alpha^5 z_2, \alpha^5 z_3) \,,\\
(x,z_1,z_2,z_3) &\sim (x, -z_1, -z_2, (-1)^2 z_3) \,,
\end{split}
\end{equation}
where $\alpha=e^{2i\pi/6}$ and $-1=e^{2i\pi/2}$. The breaking described in the first line of eq. \ref{eq:decomp} is achieved by the boundary condition given by $q_{X'}/6$. It is further broken to the second line by the boundary condition $q_{T_8^F}/2$.  The third line is achieved by further appyling $q_{Y}/2$. The above conditions enable us 
to decompose the $\textbf{248}$ representation of $E_8$ into the representations of unbroken $G_{PS}$ symmetry summarised in Table~\ref{tab:pps2}, together with the corresponding charges \footnote{The orbifold rotational conditions in eq. \ref{eq:orbcondi}, are slightly different from the previous work in \cite{Aranda:2020noz}, which preserved the flavor symmetry $SU(3)_F$. In this work the boundary conditions only preserve $SU(2)_F\times U(1)_F$ but, as will be seen below, these allow the Wilson lines to completely break the remaining symmetry into the SM one. This wasn't possible in the previous setup, making this one preferable.} (for a better presentation we color code the representations containing \textcolor{blue}{SM fermions}, \textcolor{violet}{right handed neutrinos},  \textcolor{OliveGreen}{SM Higgses}, \textcolor{magenta}{gauge fields in the adjoint}, \textcolor{red}{mirror fermions}, \textcolor{orange}{mirror Higgses}, \textcolor{brown}{flavons}, leptoquarks and vector-like triplets).
\begin{table}
	\centering
	\scriptsize
	\renewcommand{\arraystretch}{1.1}
	\begin{tabular}[t]{l|llll}
		\hline
		 & $V$ & $\phi_1$ & $\phi_2$ & $\phi_3$\\ 
		\hline
	$\mathcal{V}_{\textcolor{magenta}{(\textbf{15},\textbf{1},\textbf{1},0,0,\textbf{1})}} $ & $1,1$ & $\alpha^2, -1$ & $\alpha^5, -1$& $\alpha^5, 1$\\
		$\mathcal{V}_{\textcolor{magenta}{(\textbf{1},\textbf{3},\textbf{1},0,0,\textbf{1})}} $ & $1,1$ & $\alpha^2, -1$ & $\alpha^5, -1$& $\alpha^5, 1$\\
		$\mathcal{V}_{\textcolor{magenta}{(\textbf{1},\textbf{1},\textbf{3},0,0,\textbf{1})}} $ & $1,1$ & $\alpha^2, -1$ & $\alpha^5, -1$& $\alpha^5, 1$\\
		$\mathcal{V}_{\textcolor{magenta}{(\textbf{1},\textbf{1},\textbf{1},0,0,\textbf{1})}} $ & $1,1$ & $\alpha^2, -1$ & $\alpha^5, -1$& $\alpha^5, 1$\\
		$\mathcal{V}_{\textcolor{magenta}{(\textbf{1},\textbf{1},\textbf{1},0,0,\textbf{3})}} $ & $1,1$ & $\alpha^2, -1$ & $\alpha^5, -1$& $\alpha^5, 1$\\
		$\mathcal{V}_{\textcolor{magenta}{(\textbf{1},\textbf{1},\textbf{1},0,0,\textbf{1})}} $ & $1,1$ & $\alpha^2, -1$ & $\alpha^5, -1$& $\alpha^5, 1$\\
		$\mathcal{V}_{\textcolor{brown}{(\textbf{1},\textbf{1},\textbf{1},0,-3,\textbf{2})}} $ & $1,-1$ & $\alpha^2, 1$ & $\alpha^5, 1$& $\alpha^5, -1$\\
		$\mathcal{V}_{\textcolor{brown}{(\textbf{1},\textbf{1},\textbf{1},3,0,\textbf{2})}} $ & $1,-1$ & $\alpha^2, 1$ & $\alpha^5, 1$& $\alpha^5, -1$\\
		$\mathcal{V}_{(\textbf{6},\textbf{2},\textbf{2},0,0,\textbf{1})} $& $1,-1$ & $\alpha^2, 1$ & $\alpha^5, 1$& $\alpha^5, -1$\\
		$\mathcal{V}_{(\textbf{4},\textbf{2},\textbf{1},-3,0,\textbf{1})} $  & $\alpha^3,-1$ & $\alpha^5, 1$ & $\alpha^2, 1$& $\alpha^2, -1$\\
		$\mathcal{V}_{(\bar{\textbf{4}},\textbf{1},\textbf{2},-3,0,\textbf{1})} $ & $\alpha^3,1$ & $\alpha^5, -1$ & $\alpha^2, -1$& $\alpha^2, 1$\\
		$\mathcal{V}_{(\bar{\textbf{4}},\textbf{2},\textbf{1},3,0,\textbf{1})} $  & $\alpha^3,-1$ & $\alpha^5, 1$ & $\alpha^2, 1$& $\alpha^2, -1$\\		
		$\mathcal{V}_{(\textbf{4},\textbf{1},\textbf{2},3,0,\textbf{1})} $ & $\alpha^3,1$ & $\alpha^5, -1$ & $\alpha^2, -1$& $\alpha^2, 1$ \\	
		$\mathcal{V}_{(\textbf{6},\textbf{1},\textbf{1},-2,1,\textbf{2})} $ & $\alpha^4,-1$ & $1, 1$ & $\alpha^3,1$& $\alpha^3, -1$\\
		$\mathcal{V}_{(\textbf{6},\textbf{1},\textbf{1},-2,-2,\textbf{1})} $ & $\alpha^4,1$ & $1, -1$ & $\alpha^3,- 1$& $\alpha^3, 1$\\
		$\mathcal{V}_{(\textbf{6},\textbf{1},\textbf{1},2,-1,\textbf{2})} $  & $\alpha^2,-1$ & $\alpha^4, 1$ & $\alpha,1$& $\alpha, -1$\\	
		$\mathcal{V}_{(\textbf{6},\textbf{1},\textbf{1},2,2,\textbf{1})} $  & $\alpha^2,1$ & $\alpha^4, -1$ & $\alpha,- 1$& $\alpha, 1$\\	
		\hline
	\end{tabular}
	\hspace*{0.3cm}
	\begin{tabular}[t]{l|llll}
		\hline
		 & $V$ & $\phi_1$ & $\phi_2$ & $\phi_3$\\ 
		\hline
	$\mathcal{V}_{\textcolor{blue}{(\textbf{4},\textbf{2},\textbf{1},1,1,\textbf{2})}} $ & $\alpha,
	1$ & $\alpha^3, -1$ & $1, -1$& $1, 1$\\
	$\mathcal{V}_{\textcolor{blue}{(\textbf{4},\textbf{2},\textbf{1},1,-2,\textbf{1})}} $ & $\alpha,
	-1$ & $\alpha^3, 1$ & $1, 1$& $1, -1$\\
		$\mathcal{V}_{\textcolor{blue}{(\bar{\textbf{4}},\textbf{1},\textbf{2},1,1,\textbf{2})}} $ & $\alpha,-1$ & $\alpha^3, 1$ & $1, 1$& $1, -1$\\
		$\mathcal{V}_{\textcolor{blue}{(\bar{\textbf{4}},\textbf{1},\textbf{2},1,-2,\textbf{1})}} $ & $\alpha,1$ & $\alpha^3, -1$ & $1, -1$& $1, 1$\\
		$\mathcal{V}_{\textcolor{OliveGreen}{(\textbf{1},\textbf{2},\textbf{2},-2,1,\textbf{2})}} $  & $\alpha^4,1$ & $1, -1$ & $\alpha^3, -1$& $\alpha^3, 1$\\
		$\mathcal{V}_{\textcolor{OliveGreen}{(\textbf{1},\textbf{2},\textbf{2},-2,-2,\textbf{1})}} $  & $\alpha^4,-1$ & $1, 1$ & $\alpha^3, 1$& $\alpha^3, -1$\\
		$\mathcal{V}_{\textcolor{brown}{(\textbf{1},\textbf{1},\textbf{1},4,1,\textbf{2})}} $ & $\alpha^4,-1$ & $1, 1$ & $\alpha^3, 1$& $\alpha^3, -1$\\
		$\mathcal{V}_{\textcolor{brown}{(\textbf{1},\textbf{1},\textbf{1},4,-2,\textbf{1})}} $ & $\alpha^4,1$ & $1, -1$ & $\alpha^3, -1$& $\alpha^3, 1$\\
		$\mathcal{V}_{\textcolor{red}{(\bar{\textbf{4}},\textbf{2},\textbf{1},-1,-1,\textbf{2})}} $& $\alpha^5,1$ & $\alpha,-1$ & $\alpha^4, -1$& $\alpha^4, 1$\\
		$\mathcal{V}_{\textcolor{red}{(\bar{\textbf{4}},\textbf{2},\textbf{1},-1,2,\textbf{1})}} $& $\alpha^5,-1$ & $\alpha,1$ & $\alpha^4, 1$& $\alpha^4, -1$\\
		$\mathcal{V}_{\textcolor{red}{(\textbf{4},\textbf{1},\textbf{2},-1,-1,\textbf{2}) }} $  & $\alpha^5,-1$ & $\alpha, 1$ & $\alpha^4, 1$& $\alpha^4, -1$\\
		$\mathcal{V}_{\textcolor{red}{(\textbf{4},\textbf{1},\textbf{2},-1,2,\textbf{1})} } $  & $\alpha^5,1$ & $\alpha, -1$ & $\alpha^4, -1$& $\alpha^4, 1$\\
		$\mathcal{V}_{\textcolor{orange}{(\textbf{1},\textbf{2},\textbf{2},2,-1,\textbf{2})}} $ & $\alpha^2,1$ & $\alpha^4, -1$ & $\alpha, -1$& $\alpha, 1$\\
		$\mathcal{V}_{\textcolor{orange}{(\textbf{1},\textbf{2},\textbf{2},2,2,\textbf{1})}} $ & $\alpha^2,-1$ & $\alpha^4, 1$ & $\alpha, 1$& $\alpha, -1$\\	
		$\mathcal{V}_{\textcolor{brown}{(\textbf{1},\textbf{1},\textbf{1},-4,-1,\textbf{2})}} $  & $\alpha^2,-1$ & $\alpha^4, 1$ & $\alpha, 1$& $\alpha, -1$\\	
		$\mathcal{V}_{\textcolor{brown}{(\textbf{1},\textbf{1},\textbf{1},-4,2,\textbf{1})}} $  & $\alpha^2,1$ & $\alpha^4, -1$ & $\alpha, -1$& $\alpha, 1$\\	
		\hline
	\end{tabular}
	\caption{The $\mathbb{Z}_6\times \mathbb{Z}_2$ orbifold charges of each $G_{PS}$ $\mathcal{N}=1$ superfield. Only the superfields with both charges equal to unity 
	(the singlets $1,1$) have zero modes. The colors indicate where the different fields of interest are contained: \textcolor{blue}{SM fermions}, \textcolor{violet}{right handed neutrinos},  \textcolor{OliveGreen}{SM Higgses}, \textcolor{magenta}{gauge fields in the adjoint}, \textcolor{red}{mirror fermions}, \textcolor{orange}{mirror Higgses}, \textcolor{brown}{flavons}, leptoquarks and vector-like triplets } 
	\label{tab:pps2}
\end{table}

The zero modes have the following representations of the residual symmetry group (recall they correspond to those with $1,1$ charges in Table~\ref{tab:pps2}):
\begin{equation} 
\begin{split}
V_\mu &: \textcolor{magenta}{(\textbf{15},\textbf{1},\textbf{1},0,0,\textbf{1})}+\textcolor{magenta}{(\textbf{1},\textbf{3},\textbf{1},0,0,\textbf{1})}+\textcolor{magenta}{(\textbf{1},\textbf{1},\textbf{3},0,0,\textbf{1})}\\
&\quad+\textcolor{magenta}{(\textbf{1},\textbf{1},\textbf{1},0,0,\textbf{1})}+\textcolor{magenta}{(\textbf{1},\textbf{1},\textbf{1},0,0,\textbf{1})}+\textcolor{magenta}{(\textbf{15},\textbf{1},\textbf{1},0,0,\textbf{3})},\\
\phi_1&: (\textbf{6},\textbf{1},\textbf{1},-2,1,\textbf{2})+\textcolor{OliveGreen}{(\textbf{1},\textbf{2},\textbf{2},-2,-2,\textbf{1})}+\textcolor{brown}{(\textbf{1},\textbf{1},\textbf{1},4,1,\textbf{2})},\\
\phi_2&: \textcolor{blue}{(\textbf{4},\textbf{2},\textbf{1},1,-2,\textbf{1})}+\textcolor{blue}{(\bar{\textbf{4}},\textbf{1},\textbf{2},1,1,\textbf{2})},\\
\phi_3&: \textcolor{blue}{(\textbf{4},\textbf{2},\textbf{1},1,1,\textbf{2})}+\textcolor{blue}{(\bar{\textbf{4}},\textbf{1},\textbf{2},1,-2,\textbf{1})}.
\label{eq:zmf}
\end{split}\end{equation}
which can be named as
\begin{equation} 
\begin{split}
V_\mu &: \textcolor{magenta}{G_\mu}+\textcolor{magenta}{W^L_\mu}+\textcolor{magenta}{W^R_\mu}+\textcolor{magenta}{Z'_\mu}+\textcolor{magenta}{Z^F_\mu}+\textcolor{magenta}{W^F_\mu},\\
\phi_1&: T+\textcolor{OliveGreen}{h}+\textcolor{brown}{\Phi},\\
\phi_2&: \textcolor{blue}{f}+\textcolor{blue}{F^c},\\
\phi_3&:\textcolor{blue}{F}+\textcolor{blue}{f^c},
\end{split}\end{equation}
where uppercase letters $T,F,\Phi$ are used to denote $SU(2)_F$ doublets and lowercase letters $f,h$ correspond to $SU(2)_F$ singlets.

In order to further break the intermediate symmetry group down to the SM, one could use each Wilson line to give an effective VEV to the SM singlet fields
\begin{equation} 
\begin{split}
\braket{\phi_1}&: \textcolor{brown}{\braket{\Phi}},\\
\braket{\phi_2}&: \textcolor{blue}{\braket{F^c}},\\
\braket{\phi_3}&:\textcolor{blue}{\braket{f^c}},
\end{split}
\label{eq:wilson}
\end{equation}
where, in principle, one could choose an arbitrary scale for each effective VEV, as they come from the arbitrary Wilson lines and not from a potential. Their natural scale would be close to the compactification scale $\Lambda$. The effective VEV $\braket{\textcolor{blue}{F^c}}=(\braket{\textcolor{violet}{\tilde{\nu}^c_2}},\braket{\textcolor{violet}{\tilde{\nu}^c_3}})$ and $\braket{\textcolor{blue}{f^c}}=\braket{\textcolor{violet}{\tilde{\nu}^c_1}}$ should be aligned in the SM singlets, i.e. the right-handed sneutrino. They are charged under $SU(4)_{PS}\times SU(2)_R\times U(1)_{X'}\times U(1)_F\times SU(2)_F$ so their VEVs  break it. However, they are neutral under a $U(1)_N$ whose charge is defined by
\begin{equation}
q_N=q_X+5q_{X'} \,,
\end{equation}
and therefore do not break it. Note that the VEV $\braket{\textcolor{brown}{\Phi}}=(\textcolor{brown}{\braket{\varphi_2}},\textcolor{brown}{\braket{\varphi_3}})$ has a non-zero $q_N$ charge and hence breaks the remaining $U(1)_N$ symmetry yielding the SM gauge symmetry at low energies. Therefore, the complete geometrical breaking, involving the orbifold boundary conditions and Wilson lines, reduces $E_8\to G_{SM}\equiv SU(3)_C\times SU(2)_L\times U(1)_Y$ as desired.

\section{Anomaly-canceling sector}
\label{sec:anom}

We have shown that the orbifold breaks the $E_8$ symmetry down to the SM one leaving the desired SM field content. However, one has to ensure that any potential anomalies
are cancelled at every symmetry breaking step. Let us discuss this point in more detail.

The rotation boundary conditions split the masses of different representations by integer multiples of the compactification scale $n\Lambda$. The Wilson lines split the masses  by an effective VEV with scale $y\Lambda$, where $y$ is an arbitrary real parameter. The latter can, in principle, be as small as desired, generating a certain hierarchy between the scales. Therefore, one could think of the subsequential breaking $E_8\to G_{PS} \to G_{SM}$ at well-separated energy scales as a New Physics scenario potentially relevant for phenomenology. 

The field content in the Pati-Salam phase would be composed of zero modes from Eq.~(\ref{eq:zmf}). However, we can easily notice that this phase is not fully consistent as the field content generates gauge anomalies, while the SM phase is anomaly free. In order to resolve this issue, one can add more fields so that the considered two-step breaking is consistently realised in an anomaly-free way.

There are two sets of fields that cancel the anomalies generated by the zero modes. First, one would add 27 copies of the pair of representations $\textcolor{brown}{(\textbf{1},\textbf{1},\textbf{1},0,-1,\textbf{2})}+\textcolor{brown}{(\textbf{1},\textbf{1},\textbf{1},0,2,\textbf{1})}$ as the 4d chiral superfields located at the $z_i=0$ origin brane. These would cancel the flavor-specific anomalies.
The second set of fields consists of 5 copies of 10d chiral superfields in the bulk living in the $\textbf{248}$ representation of $E_8$. They have the charges under the orbifold rotations $(1,\pm 1),(\alpha^4,\pm 1), (\alpha^2,-1)$. These will add a bunch of zero-mode chiral superfields (51 in total) which are either the real representations (i.e. vector-like pairs) or the chiral representations ${(\textbf{6},\textbf{1},\textbf{1},-2,-2,\textbf{1})}+\textcolor{OliveGreen}{(\textbf{1},\textbf{2},\textbf{2},-2,-2,\textbf{1})}+\textcolor{brown}{(\textbf{1},\textbf{1},\textbf{1},4,-2,\textbf{1})}$, as summarised in Table~\ref{tab:pps5}.

The SM fields come from the gauge superfield. While one can add a symmetry to restrict the couplings, they will not change the couplings to the gauge superfield. This does not reduce the predictivity of the setup. The added fields will have arbitrary couplings and, since they live in real representations, explicit arbitrary masses.
\begin{table}
	\centering
	\tiny
	\renewcommand{\arraystretch}{1}
	\begin{tabular}[t]{l|llll}
		\hline
		 & $\phi_0$ & $\phi_1$ & $\phi_2$ & $\phi_3$\\ 
		\hline
	$\Phi_{\textcolor{magenta}{(\textbf{15},\textbf{1},\textbf{1},0,0,\textbf{1})}} $ & \colorbox{Cerulean}{$1,1$} & \colorbox{Apricot}{$\alpha^2, -1$} & $\alpha^5, -1$& $\alpha^5, 1$\\
		$\Phi_{\textcolor{magenta}{(\textbf{1},\textbf{3},\textbf{1},0,0,\textbf{1})}} $ & \colorbox{Cerulean}{$1,1$} & \colorbox{Apricot}{$\alpha^2, -1$} & $\alpha^5, -1$& $\alpha^5, 1$\\
		$\Phi_{\textcolor{magenta}{(\textbf{1},\textbf{1},\textbf{3},0,0,\textbf{1})}} $ & \colorbox{Cerulean}{$1,1$} & \colorbox{Apricot}{$\alpha^2, -1$} & $\alpha^5, -1$& $\alpha^5, 1$\\
		$\Phi_{\textcolor{magenta}{(\textbf{1},\textbf{1},\textbf{1},0,0,\textbf{1})}} $ & \colorbox{Cerulean}{$1,1$} & \colorbox{Apricot}{$\alpha^2, -1$} & $\alpha^5, -1$& $\alpha^5, 1$\\
		$\Phi_{\textcolor{magenta}{(\textbf{1},\textbf{1},\textbf{1},0,0,\textbf{3})}} $ & \colorbox{Cerulean}{$1,1$} & \colorbox{Apricot}{$\alpha^2, -1$} & $\alpha^5, -1$& $\alpha^5, 1$\\
		$\Phi_{\textcolor{magenta}{(\textbf{1},\textbf{1},\textbf{1},0,0,\textbf{1})}} $ & \colorbox{Cerulean}{$1,1$} & \colorbox{Apricot}{$\alpha^2, -1$} & $\alpha^5, -1$& $\alpha^5, 1$\\
		$\Phi_{\textcolor{brown}{(\textbf{1},\textbf{1},\textbf{1},0,-3,\textbf{2})}} $ & \colorbox{LimeGreen}{$1,-1$} & $\alpha^2, 1$ & $\alpha^5, 1$& $\alpha^5, -1$\\
		$\Phi_{\textcolor{brown}{(\textbf{1},\textbf{1},\textbf{1},0,3,\textbf{2})}} $ & \colorbox{LimeGreen}{$1,-1$} & $\alpha^2, 1$ & $\alpha^5, 1$& $\alpha^5, -1$\\
		$\Phi_{(\textbf{6},\textbf{2},\textbf{2},0,0,\textbf{1})} $& \colorbox{LimeGreen}{$1,-1$} & $\alpha^2, 1$ & $\alpha^5, 1$& $\alpha^5, -1$\\
		$\Phi_{(\textbf{4},\textbf{2},\textbf{1},-3,0,\textbf{1})} $  & $\alpha^3,-1$ & $\alpha^5, 1$ & $\alpha^2, 1$& $\alpha^2, -1$\\
		$\Phi_{(\bar{\textbf{4}},\textbf{1},\textbf{2},-3,0,\textbf{1})} $ & $\alpha^3,1$ & $\alpha^5, -1$ & $\alpha^2, -1$& $\alpha^2, 1$\\
		$\Phi_{(\bar{\textbf{4}},\textbf{2},\textbf{1},3,0,\textbf{1})} $  & $\alpha^3,-1$ & $\alpha^5, 1$ & $\alpha^2, 1$& $\alpha^2, -1$\\		
		$\Phi_{(\textbf{4},\textbf{1},\textbf{2},3,0,\textbf{1})} $ & $\alpha^3,1$ & $\alpha^5, -1$ & $\alpha^2, -1$& $\alpha^2, 1$ \\	
		$\Phi_{(\textbf{6},\textbf{1},\textbf{1},-2,1,\textbf{2})} $ & \colorbox{Thistle}{$\alpha^4,-1$} & \colorbox{Cerulean}{$1, 1$} & $\alpha^3,1$& $\alpha^3, -1$\\
		$\Phi_{(\textbf{6},\textbf{1},\textbf{1},-2,-2,\textbf{1})} $ & \colorbox{Goldenrod}{$\alpha^4,1$} & \colorbox{LimeGreen}{$1, -1$} & $\alpha^3,- 1$& $\alpha^3, 1$\\
		$\Phi_{(\textbf{6},\textbf{1},\textbf{1},2,-1,\textbf{2})} $  & \colorbox{Apricot}{$\alpha^2,-1$} & \colorbox{Goldenrod}{$\alpha^4, 1$} & $\alpha,1$& $\alpha, -1$\\	
		$\Phi_{(\textbf{6},\textbf{1},\textbf{1},2,2,\textbf{1})} $  & $\alpha^2,1$ & \colorbox{Thistle}{$\alpha^4, -1$} & $\alpha,- 1$& $\alpha, 1$\\	
		\hline
	\end{tabular}
	\hspace*{0.2cm}
	\begin{tabular}[t]{l|llll}
		\hline
		 & $\phi_0$ & $\phi_1$ & $\phi_2$ & $\phi_3$\\ 
		\hline
	$\Phi_{\textcolor{blue}{(\textbf{4},\textbf{2},\textbf{1},1,1,\textbf{2})}} $ & $\alpha,
	1$ & $\alpha^3, -1$ & \colorbox{LimeGreen}{$1, -1$}& \colorbox{Cerulean}{$1, 1$}\\
	$\Phi_{\textcolor{blue}{(\textbf{4},\textbf{2},\textbf{1},1,-2,\textbf{1})}} $ & $\alpha,
	-1$ & $\alpha^3, 1$ & \colorbox{Cerulean}{$1, 1$}& \colorbox{LimeGreen}{$1, -1$}\\
		$\Phi_{\textcolor{blue}{(\bar{\textbf{4}},\textbf{1},\textbf{2},1,1,\textbf{2})}} $ & $\alpha,-1$ & $\alpha^3, 1$ & \colorbox{Cerulean}{$1, 1$}& \colorbox{LimeGreen}{$1, -1$}\\
		$\Phi_{\textcolor{blue}{(\bar{\textbf{4}},\textbf{1},\textbf{2},1,-2,\textbf{1})}} $ & $\alpha,1$ & $\alpha^3, -1$ & \colorbox{LimeGreen}{$1, -1$}& \colorbox{Cerulean}{$1, 1$}\\
		$\Phi_{\textcolor{OliveGreen}{(\textbf{1},\textbf{2},\textbf{2},-2,1,\textbf{2})}} $  & \colorbox{Goldenrod}{$\alpha^4,1$} & \colorbox{LimeGreen}{$1, -1$} & $\alpha^3, -1$& $\alpha^3, 1$\\
		$\Phi_{\textcolor{OliveGreen}{(\textbf{1},\textbf{2},\textbf{2},-2,-2,\textbf{1})}} $  & \colorbox{Thistle}{$\alpha^4,-1$} & \colorbox{Cerulean}{$1, 1$} & $\alpha^3, 1$& $\alpha^3, -1$\\
		$\Phi_{\textcolor{brown}{(\textbf{1},\textbf{1},\textbf{1},4,1,\textbf{2})}} $ & \colorbox{Thistle}{$\alpha^4,-1$} & \colorbox{Cerulean}{$1, 1$} & $\alpha^3, 1$& $\alpha^3, -1$\\
		$\Phi_{\textcolor{brown}{(\textbf{1},\textbf{1},\textbf{1},4,-2,\textbf{1})}} $ & \colorbox{Goldenrod}{$\alpha^4,1$} & \colorbox{LimeGreen}{$1, -1$ }& $\alpha^3, -1$& $\alpha^3, 1$\\
		$\Phi_{\textcolor{red}{(\bar{\textbf{4}},\textbf{2},\textbf{1},-1,-1,\textbf{2})}} $& $\alpha^5,1$ & $\alpha,-1$ & \colorbox{Thistle}{$\alpha^4, -1$}& \colorbox{Goldenrod}{$\alpha^4, 1$}\\
		$\Phi_{\textcolor{red}{(\bar{\textbf{4}},\textbf{2},\textbf{1},-1,2,\textbf{1})}} $& $\alpha^5,-1$ & $\alpha,1$ & \colorbox{Goldenrod}{$\alpha^4, 1$}& \colorbox{Thistle}{$\alpha^4, -1$}\\
		$\Phi_{\textcolor{red}{(\textbf{4},\textbf{1},\textbf{2},-1,-1,\textbf{2}) }} $  & $\alpha^5,-1$ & $\alpha, 1$ & \colorbox{Goldenrod}{$\alpha^4, 1$}& \colorbox{Thistle}{$\alpha^4, -1$}\\
		$\Phi_{\textcolor{red}{(\textbf{4},\textbf{1},\textbf{2},-1,2,\textbf{1})} } $  & $\alpha^5,1$ & $\alpha, -1$ & \colorbox{Thistle}{$\alpha^4, -1$}& \colorbox{Goldenrod}{$\alpha^4, 1$}\\
		$\Phi_{\textcolor{orange}{(\textbf{1},\textbf{2},\textbf{2},2,-1,\textbf{2})}} $ & $\alpha^2,1$ & \colorbox{Thistle}{$\alpha^4, -1$} & $\alpha, -1$& $\alpha, 1$\\
		$\Phi_{\textcolor{orange}{(\textbf{1},\textbf{2},\textbf{2},2,2,\textbf{1})}} $ & \colorbox{Apricot}{$\alpha^2,-1$} & \colorbox{Goldenrod}{$\alpha^4, 1$} & $\alpha, 1$& $\alpha, -1$\\	
		$\Phi_{\textcolor{brown}{(\textbf{1},\textbf{1},\textbf{1},-4,-1,\textbf{2})}} $  & \colorbox{Apricot}{$\alpha^2,-1$} & \colorbox{Goldenrod}{$\alpha^4, 1$} & $\alpha, 1$& $\alpha, -1$\\	
		$\Phi_{\textcolor{brown}{(\textbf{1},\textbf{1},\textbf{1},-4,2,\textbf{1})}} $  & $\alpha^2,1$ & \colorbox{Thistle}{$\alpha^4, -1$} & $\alpha, -1$& $\alpha, 1$\\	
		\hline
	\end{tabular}
	\caption{The $\mathbb{Z}_6\times \mathbb{Z}_2$ orbifold charges of each $G_{PS}$ $\mathcal{N}=1$ superfield. Only the superfields with highlighted charges have zero modes from the 10d chiral superfields. Same colored highlighted fields would come from the same 10d chiral superfield.} 
	\label{tab:pps5}
\end{table}

Below the compactification scale we have the chiral multiplets
\begin{equation}
\begin{array}{llll}
2\times\Phi_{\textcolor{magenta}{(\textbf{15},\textbf{1},\textbf{1},0,0,\textbf{1})}}, &   0\times\Phi_{(\textbf{6},\textbf{2},\textbf{2},0,0,\textbf{1})} ,
& 3\times \Phi_{\textcolor{blue}{(\textbf{4},\textbf{2},\textbf{1},1,1,\textbf{2})}} ,
&2\times\Phi_{\textcolor{red}{(\bar{\textbf{4}},\textbf{2},\textbf{1},-1,-1,\textbf{2})}},
\\
2\times \Phi_{\textcolor{magenta}{(\textbf{1},\textbf{3},\textbf{1},0,0,\textbf{1})}},
& 0\times \Phi_{(\textbf{4},\textbf{2},\textbf{1},-3,0,\textbf{1})} ,
& 3 \times \Phi_{\textcolor{blue}{(\textbf{4},\textbf{2},\textbf{1},1,-2,\textbf{1})}} ,
&2 \times \Phi_{\textcolor{red}{(\bar{\textbf{4}},\textbf{2},\textbf{1},-1,2,\textbf{1})}} ,
 \\
2\times\Phi_{\textcolor{magenta}{(\textbf{1},\textbf{1},\textbf{3},0,0,\textbf{1})}} ,
&0\times \Phi_{(\bar{\textbf{4}},\textbf{1},\textbf{2},-3,0,\textbf{1})},
& 3 \times \Phi_{\textcolor{blue}{(\bar{\textbf{4}},\textbf{1},\textbf{2},1,1,\textbf{2})}},
& 2\times\Phi_{\textcolor{red}{(\textbf{4},\textbf{1},\textbf{2},-1,-1,\textbf{2}) }},
\\
2\times \Phi_{\textcolor{magenta}{(\textbf{1},\textbf{1},\textbf{1},0,0,\textbf{1})}},
& 0\times \Phi_{(\bar{\textbf{4}},\textbf{2},\textbf{1},3,0,\textbf{1})} ,
& 3 \times \Phi_{\textcolor{blue}{(\bar{\textbf{4}},\textbf{1},\textbf{2},1,-2,\textbf{1})}} ,
&2\times \Phi_{\textcolor{red}{(\textbf{4},\textbf{1},\textbf{2},-1,2,\textbf{1})} },
 \\
2\times \Phi_{\textcolor{magenta}{(\textbf{1},\textbf{1},\textbf{1},0,0,\textbf{3})}},
&0\times \Phi_{(\textbf{4},\textbf{1},\textbf{2},3,0,\textbf{1})},
& 2 \times \Phi_{\textcolor{OliveGreen}{(\textbf{1},\textbf{2},\textbf{2},-2,1,\textbf{2})}},
 & 1\times \Phi_{\textcolor{orange}{(\textbf{1},\textbf{2},\textbf{2},2,-1,\textbf{2})}} ,
 \\
2\times \Phi_{\textcolor{magenta}{(\textbf{1},\textbf{1},\textbf{1},0,0,\textbf{1})}},
&  3\times \Phi_{(\textbf{6},\textbf{1},\textbf{1},-2,1,\textbf{2})} ,
& 3 \times \Phi_{\textcolor{OliveGreen}{(\textbf{1},\textbf{2},\textbf{2},-2,-2,\textbf{1})}},
& 2\times \Phi_{\textcolor{orange}{(\textbf{1},\textbf{2},\textbf{2},2,2,\textbf{1})}} ,
\\
 1\times \Phi_{\textcolor{brown}{(\textbf{1},\textbf{1},\textbf{1},0,-3,\textbf{2})}} ,
 &2 \times \Phi_{(\textbf{6},\textbf{1},\textbf{1},-2,-2,\textbf{1})},
 & 3 \times \Phi_{\textcolor{brown}{(\textbf{1},\textbf{1},\textbf{1},4,1,\textbf{2})}} ,
 & 2 \times \Phi_{\textcolor{brown}{(\textbf{1},\textbf{1},\textbf{1},-4,-1,\textbf{2})}},
 \\
 1\times \Phi_{\textcolor{brown}{(\textbf{1},\textbf{1},\textbf{1},0,3,\textbf{2})}},
 & 2\times \Phi_{(\textbf{6},\textbf{1},\textbf{1},2,-1,\textbf{2})},
 & 2 \times \Phi_{\textcolor{brown}{(\textbf{1},\textbf{1},\textbf{1},4,-2,\textbf{1})}},
 & 1 \times \Phi_{\textcolor{brown}{(\textbf{1},\textbf{1},\textbf{1},-4,2,\textbf{1})}} ,
 \\
  &  1\times \Phi_{(\textbf{6},\textbf{1},\textbf{1},2,2,\textbf{1})} ,
  & 27\times \Phi_{\textcolor{brown}{(\textbf{1},\textbf{1},\textbf{1},0,-1,\textbf{2})} }
  &27 \times \Phi_{\textcolor{brown}{(\textbf{1},\textbf{1},\textbf{1},0,2,\textbf{1})}}
\label{eq:extrafields}
 \end{array}
 \end{equation}
 where all the vector-like pairs have an arbitrary mass determined by the above mentioned parameters. The unpaired fields contain the SM sectors plus SM singlets as we further discuss below in \cref{sec:RGE}. While the masses of the KK modes are fixed by the compactification scale, the ones in \cref{eq:extrafields} are determined  by the arbitrary parameters of the superpotential. Furthermore, a Wilson line VEV does not preserve SUSY, therefore the chiral multiplets will be split into their scalar and fermion components.
 
 For completeness, let us also show the branching rules of the Pati-Salam {\blue blue} representations in terms of the SM gauge group identifying them with standard chiral matter and right-handed neutrinos,
 \begin{equation}
     \begin{aligned}
     &{\color{blue} (\bm{4},\bm{2},\bm{1},1,1,\bm{2}) } \to {\color{blue}(\bm{3},\bm{2},-\dfrac{1}{6})} + {\blue {\color{blue}(\bm{1},\bm{2},\dfrac{1}{2})}} \equiv { \blue\mathrm{Q}_{1,2} + \mathrm{L}_{1,2} }
     \\
     &{\color{blue} (\bm{4},\bm{2},\bm{1},1,-2,\bm{1}) } \to {\color{blue}(\bm{3},\bm{2},-\dfrac{1}{6})} + {\blue {\color{blue}(\bm{1},\bm{2},\dfrac{1}{2})}} \equiv { \blue\mathrm{Q}_3 + \mathrm{L}_{3} }
     \\
     &{\color{blue} (\bar{\textbf{4}},\textbf{1},\textbf{2},1,1,\textbf{2}) } \to {\blue (\bm{\overline{3}},\bm{1},-\frac{2}{3})} + {\color{blue}(\bm{\overline{3}},\bm{1},\frac{1}{3})} + {\blue {\color{blue}(\bm{1},\bm{1},1)}} + {\blue (\bm{1},\bm{1},0) } \equiv { \blue \mathrm{u}^c_{1,2} + \mathrm{d}^c_{1,2} + \mathrm{e}^c_{1,2} + \mathrm{\nu}^c_{1,2} }
     \\
     &{\color{blue} (\bar{\textbf{4}},\textbf{1},\textbf{2},1,-2,\textbf{1}) } \to {\blue (\bm{\overline{3}},\bm{1},-\frac{2}{3})} + {\color{blue}(\bm{\overline{3}},\bm{1},\frac{1}{3})} + {\blue {\color{blue}(\bm{1},\bm{1},1)}} + {\blue (\bm{1},\bm{1},0) } \equiv { \blue \mathrm{u}^c_{3} + \mathrm{d}^c_{3} + \mathrm{e}^c_{3} + \mathrm{\nu}^c_{3} }
     \end{aligned}
     \label{eq:SMid}
 \end{equation}
 with the labels $1,2,3$ denoting the three families.

\section{RG evolution of gauge couplings}
\label{sec:RGE}

In this section, we study the RG evolution of the gauge couplings in our model with the purpose of finding possible low-energy scale scenarios compatible with an exact unification of all forces, including flavour, into $\mathrm{E}_8$. Our strategy for the current analysis consists in searching for those extensions of the SM with a minimal field content. Note that for a consistent SM-like fermion mass spectrum one needs two Higgs doublets. While one is responsible for giving masses to up-type quarks the other is needed for their down-type partners. Therefore, the minimal framework to consider contains two Higgs doublets, commonly dubbed two-Higgs doublet model (2HDM) (for a detailed review, see e.g.~Refs.~\cite{Branco:1999fs,Branco:2011iw,Ivanov:2017dad}).

In what follows we study whether a 2HDM scenario is already consistent with gauge couplings unification under $\mathrm{E}_8$ and, if not, how many extra Higgs doublets or generations of vector-like fermions are needed to fulfil the unification condition. As we discuss below, we will only consider low-scale scenarios with up to a maximum of three generations of vector-like quarks (VLQ) and vector-like leptons (VLL) that can be either $\SU{2}{L}$ doublets or singlets. We also allow up to one additional Higgs doublet in the low-energy scale theory on top of the two doublets already mentioned above.

The running of the gauge couplings is calculated at one-loop order where the value of the inverse structure constants at a given scale $\mu$ is given by
\begin{equation}
\alpha_\mathrm{A}^{-1}\left(\mu{}{}\right) = \alpha_0^{-1} + \frac{b_\mathrm{A}}{2 \pi} \log\left(\frac{\mu}{\mu_0}\right) \,,
\label{eq:RGE}
\end{equation}
where a label $\mathrm{A}$ identifies a given gauge group and $\alpha_\mathrm{A} = g_\mathrm{A}^2/(4 \pi)$, while $\alpha_0^{-1}$ denotes the value of the inverse structure constant at the initial energy-scale $\mu_0$. The value of the $b_\mathrm{A}$ coefficients will determine how fast a given gauge coupling evolves between any two scales. For non-Abelian gauge groups these are given by
\begin{equation}
    b_\mathrm{A} = \frac{11}{3} C_2\left(G\right) - \frac{4}{3} \kappa T\left(F\right) - \frac{1}{3} T\left(S\right)\,,
    \label{eq:SUrun}
\end{equation}
where $\kappa = \tfrac{1}{2}$ for Weyl fermions, $C_2\left(G\right)$ is a group Casimir in the adjoint representation, $T\left(F\right)$ and $T\left(S\right)$ are the Dynkin indices for fermions and complex scalars, respectively. For the case of $\U{}$ symmetries the beta-function coefficients read as
\begin{equation}
    b^\prime_\mathrm{A} = -\frac{4}{3} \kappa \sum_f \left(\frac{Q_f}{2}\right)^2 - \frac{1}{3} \sum_s \left(\frac{Q_s}{2}\right)^2 \,,
    \label{eq:U1run}
\end{equation}
with $Q_f$ and $Q_s$ the Abelian charges of the fermions and scalars in the theory.

In our RG analysis, we consider three distinct regions:
\begin{enumerate}
    \item The first one corresponds to the running of the gauge couplings of the $G_{PS}$ symmetry emergent from the orbifold compactification of the 10-dimensional $\mathrm{E}_8$ theory as described above. In such a region we label the beta-function coefficients as $b^\mathrm{I}_\mathrm{A}$ and denote the universal $\mathrm{E}_8$ inverse structure constant at the GUT scale as $\alpha_8^{-1}$. At this stage, all representations identified in \cref{eq:extrafields} contribute to the $b^\mathrm{I}_\mathrm{A}$ coefficients with the indicated multiplicity. Knowing that for $\SU{4}{}$ we have
    \begin{equation}
        T\left(\bm{15}\right) = C_2\left(\bm{15}\right) = 4\,, \qquad
        T\left(\bm{6}\right) = 1\,, \qquad
        T\left(\bm{4}\right) = \frac12\,,
    \end{equation}
    and that for $\SU{2}{}$
    \begin{equation}
        T\left(\bm{3}\right) = C_2\left(\bm{3}\right) = 2\,, \qquad
        T\left(\bm{2}\right) = \frac12\,,
    \end{equation}
    we obtain the coefficients
    \begin{equation}
        b^\mathrm{I}_\mathrm{PS} = -40\,, \qquad b^\mathrm{I}_\mathrm{L} = b^\mathrm{I}_\mathrm{R} = -47\,, \qquad
        b^\mathrm{I}_\mathrm{F} = -78\,.
    \end{equation}
    Taking into account the $\U{X'}$ and $\U{F}$ charges and respective multiplicities also in \cref{eq:extrafields}, the slopes of the RGEs of the Abelian inverse structure constants read as
    \begin{equation}
        b^{' \mathrm{I}}_\mathrm{X'} = -216\,, \qquad b^{' \mathrm{I}}_\mathrm{F} = -279\,.
    \end{equation}
    \item The second region corresponds to the stage after the three Wilson lines $\braket{\phi_{1,2,3}}$ give VEVs to the SM singlet directions as specified in \cref{eq:wilson}. The gauge group after this stage is that of the SM and, as discussed above, we only study the following possibilities:
    \begin{itemize}
        \item A scalar sector with either two or three Higgs doublets that we denote as $\mathrm{H}$ in what follows;
        \item New exotic quarks containing either none or up to three generations of $\SU{2}{L}$ doublet VLQs denoted as $\mathrm{Q}_\mathrm{V}$;
        \item New exotic up-type quarks containing either none or up to three generations of $\SU{2}{L}$ singlet VLQs and denoted as $\mathrm{U}_\mathrm{V}$;
        \item New exotic down-type quarks containing either none or up to three generations of $\SU{2}{L}$ singlet VLQs and denoted as $\mathrm{D}_\mathrm{V}$;
        \item New exotic leptons containing either none or up to three generations of $\SU{2}{L}$ doublet VLLs denoted as $\mathrm{L}_\mathrm{V}$;
        \item New exotic leptons containing either none or up to three generations of $\SU{2}{L}$ singlet VLLs denoted as $\mathrm{E}_\mathrm{V}$.
    \end{itemize}
    Note that the choice of including up to three generations of vector-like fermions in the low-energy spectrum is not arbitrary. To see this let us consider the possible bilinear terms involving the {\red red} and {\blue blue} fields in \cref{eq:extrafields} that can be cast as
    \begin{equation}
        \begin{aligned}
        \sum_{i=1}^3 \sum_{k=1}^2 & \left( \mu_{ik} \textcolor{blue}{F_i} \textcolor{red}{\overline{F}_k}
        +
        \mu_{ik}^{\prime}\textcolor{blue}{f_i} \textcolor{red}{\overline{f}_k}
        +
        \mu_{ik}^{\prime \prime}\textcolor{blue}{F_i^c} \textcolor{red}{\overline{F}_k^c}
        +
        \mu_{ik}^{\prime \prime \prime}\textcolor{blue}{f_i^c} \textcolor{red}{\overline{f}_k^c}
        \right)\,.
        \end{aligned}
        \label{eq:bil}
    \end{equation}
    If we specialize on the first term we can write a $5 \times 5$ mass matrix in the basis $\left\{ \textcolor{blue}{F_1}, \textcolor{blue}{F_2}, \textcolor{blue}{F_3}, \textcolor{red}{\overline{F}_1}, \textcolor{red}{\overline{F}_2} \right\}$ as
    \begin{equation}
        \bm{m}_\mathrm{F} = 
\begin{pmatrix}
0 &0 &0 & \mu_{11} &\mu_{12} \\ 
0 &0 & 0 &\mu_{21} &\mu_{22} \\ 
0 &0 &0 &\mu_{31} &\mu_{32} \\ 
\mu_{11} &\mu_{21} &\mu_{31} &0 &0 \\ 
\mu_{12} &\mu_{22} &\mu_{32} &0 &0 
\end{pmatrix} \,,
\label{eq:mF}
    \end{equation}
    whose rank is 4. Furthermore, and assuming $\mu_{ik}$ real for illustration purposes, we see that $m_\mathrm{F}^2 = m_\mathrm{F} \cdot m_\mathrm{F}^\dagger$ has two degenerate eigenvalues which means that in the mass basis we end up with one massless chiral fermion $F$ as well as two vector-like states $F_\mathrm{V}^1$ and $F_\mathrm{V}^2$. Inspired by the unified origin of our model under $\mathrm{E}_8$, in the limit where all $\mu_{ik}$ can be thought as approximately degenerate one can write
    \begin{equation}
        \mu_{ik} = \mu + \epsilon m_{ik} \,,
    \end{equation}
    where, for $\epsilon \ll 1$ the squared masses read as
    \begin{equation}
        m^2_F = 0 \qquad m^2_{F_\mathrm{V}^1} = 0 + \mathcal{O}\left( \epsilon^2 \right) \qquad m^2_{F_\mathrm{V}^2} = 6 \mu^2 + \mathcal{O}\left( \epsilon \right)\,.
    \end{equation}
    This implies that, for a $\mu$ of the order of the compactification scale and for sufficiently small $\epsilon$, we can have $m_{F_\mathrm{V}^1} \ll m_{F_\mathrm{V}^2}$. In turn, it may result in up to two generations of vector-like fermions of the type $F_\mathrm{V}^1$ not far from the TeV scale. Note that both $\textcolor{blue}{F_i}$ and $\textcolor{red}{\overline{F}_k}$ are $\SU{2}{F}$ doublets and so is $F_\mathrm{V}^1$. The exact same reasoning can be applied to the $\mu_{ik}^\prime$, $\mu_{ik}^{\prime \prime}$ and $\mu_{ik}^{\prime \prime \prime}$ yielding up to two generations of the Pati-Salam fermions $F_\mathrm{V}^{1\,c}$ and up to one generation of $f_\mathrm{V}^{1}$ and $f_\mathrm{V}^{1\,c}$, motivating our choices in the bullet points above. Note that the $\SU{2}{L}$ doublet VLQs, $\mathrm{Q_V}$, and VLLs $\mathrm{L_V}$ belong to $F_\mathrm{V}$ (two generations) and $f_\mathrm{V}$ (one generation), while their singlet counterparts, $\mathrm{U_V}$, $\mathrm{D_V}$ and $\mathrm{E_V}$ are embedded in $F_\mathrm{V}^c$ and $f_\mathrm{V}^c$. Similarly, all chiral matter belongs to the massless eigenstates and transforms according to the {\blue blue} quantum numbers in \cref{eq:extrafields}.
    
    With this in mind, the coefficients of the $G_{SM}$ RGEs read as
    \begin{equation}
        \begin{aligned}
        b_\mathrm{C}^\mathrm{II} &= 11 -\frac23 \left( 6 + n_\mathrm{Q_V} + \frac{n_\mathrm{D_V}}{2} + \frac{n_\mathrm{U_V}}{2} \right)\,, \\
        b_\mathrm{L}^\mathrm{II} &= \frac{22}{3} - \frac13 \left( \frac{n_\mathrm{H}}{2} + 1  \right) - \frac23 \left( 6 + \frac{n_\mathrm{L_V}}{2} + 3 \frac{n_\mathrm{Q_V}}{2} \right)\,,
        \\
        b_\mathrm{Y}^\mathrm{II} &= -\frac13 \left( \frac{11}{4} + \frac{n_\mathrm{H}}{16} \right) - \frac23 \left( \frac{11}{4} + \frac{n_\mathrm{D_V}}{36} + \frac{n_\mathrm{E_V}}{4} + \frac{n_\mathrm{L_V}}{16} + \frac{n_\mathrm{Q_V}}{144} + \frac{n_\mathrm{U_V}}{9} \right)\,,
        \end{aligned}
        \label{eq:GSM-II}
    \end{equation}
    with the various $n_\mathrm{X}$, encoding the number of extra Higgs and vector-like fermions at the low scale according to the label $\mathrm{X} = \mathrm{H,Q_V,U_V,D_V,L_V,E_V}$.
    
    \item Finally, we consider a third region below the mass threshold of the vector-like fermions and where the only New Physics states are either one or two additional Higgs doublets, i.e., a 2HDM or a 3HDM EW-scale theory. Note that the presence of three Higgs doublets can be advantageous for the generation of a realistic CKM mixing in the quark sector as discussed in Ref.~\cite{Morais:2020ypd,Morais:2020odg}. With this in mind, the beta-function coefficients in this region are
    \begin{equation}
        \begin{aligned}
         b_\mathrm{C}^\mathrm{III} &= 7\,,
         \\
         b_\mathrm{L}^\mathrm{III} &= \frac{10}{3} - \frac13 \left( \frac{n_\mathrm{H}}{2} + 1  \right)\,,
         \\
         b_\mathrm{Y}^\mathrm{III} &= -\frac{11}{6} - \frac13 \left( \frac{n_\mathrm{H}}{16} + \frac{11}{4}  \right)\,.
        \end{aligned}
        \label{eq:GSM-III}
    \end{equation}
\end{enumerate}
The $\U{Y}$ generators and the inverse structure constants matched at tree-level with those of the Pati-Salam theory read as
    \begin{equation}
        \text{T}_{\text{Y}}=\sqrt{\frac{2}{3}}\text{T}_{\text{PS}}^{15}-\text{T}_{\text{R}}^{3} \qquad \alpha_\text{Y}^{-1}=\text{\ensuremath{\dfrac{2}{3}\alpha_\text{PS}^{-1}+\alpha_\text{R}^{-1}}}\,,
        \label{eq:U1match}
    \end{equation}
    resulting in the RGEs
    \begin{equation}
    \begin{aligned}
        \alpha_\text{Y}^{-1}(\mu) &= \frac{2}{3}\alpha_\text{PS}^{-1}(\mu_8)+\alpha_\text{R}^{-1}(\mu_8) + \left(\frac{b_{\text{PS}}^{\text{I}}}{3\pi}+\frac{b_{\text{R}}^{\text{I}}}{2\pi}\right)\log\left( \frac{\mu_{_\mathrm{PS}}}{\mu_8} \right) + \frac{b_{\text{Y}}^{\text{II}}}{2\pi}\log\left( \frac{\mu_{_\mathrm{VLF}}}{\mu_{_\mathrm{PS}}} \right)
        \\
        & + \frac{b_{\text{Y}}^{\text{III}}}{2\pi}\log\left( \frac{\mu}{\mu_{_\mathrm{VLF}}} \right)\,,
        \\
        \alpha_\text{L}^{-1}(\mu) &= \alpha_\text{L}^{-1}(\mu_8) + \frac{b_{\text{L}}^{\text{I}}}{2\pi}\log\left( \frac{\mu_{_\mathrm{PS}}}{\mu_8} \right) + \frac{b_{\text{L}}^{\text{II}}}{2\pi}\log\left( \frac{\mu_{_\mathrm{VLF}}}{\mu_{_\mathrm{PS}}}  \right)
        + \frac{b_{\text{L}}^{\text{III}}}{2\pi}\log\left( \frac{\mu}{\mu_{_\mathrm{VLF}}} \right)\,,
        \\
        \alpha_\text{C}^{-1}(\mu) &= \alpha_\text{PS}^{-1}(\mu_8) + \frac{b_{\text{PS}}^{\text{I}}}{2\pi}\log\left( \frac{\mu_{_\mathrm{PS}}}{\mu_8} \right) + \frac{b_{\text{C}}^{\text{II}}}{2\pi}\log\left( \frac{\mu}{\mu_{_\mathrm{VLF}}} \right)
        + \frac{b_{\text{C}}^{\text{III}}}{2\pi}\log\left( \frac{\mu}{\mu_{_\mathrm{VLF}}} \right)\,,
        \end{aligned}
        \label{eq:RGEs}
    \end{equation}
    where $\mu_8\equiv \Lambda$ is the orbifold compactification scale, $\mu_{_\mathrm{PS}}$ represents the Pati-Salam breaking scale via the Wilson line mechanism whereas $\mu_{_\mathrm{VLF}}$ is the mass threshold scale at which all vector-like fermions are integrated out.

We have performed a scan over the number of vector-like fermion generations such that $0 \leq n_\mathrm{X} \leq 3$ for $\mathrm{X} = \mathrm{Q_V,U_V,D_V,L_V,E_V}$ and additional Higgs doublets $0 \leq n_\mathrm{H} \leq 1$, i.e.~considering a scalar $(n_\mathrm{H}+2)$HDM sector.  In addition, we have selected two possible cases: one where $\mu_{_\mathrm{VLF}} = 1~\mathrm{TeV}$ and so the new VLLs and VLQs may be at the reach of the LHC (see Ref.~\cite{Freitas:2020ttd} for a recent study on VLLs phenomenology in the GUT context), and another where $\mu_{_\mathrm{VLF}} = 10~\mathrm{TeV}$, which may only be accessible at future colliders. Defining as valid low-scale models (we call each set of $(n_\mathrm{X} , n_H)$ a "model") those that are compatible with an exact unification of all gauge interactions at the $\mathrm{E}_8$ breaking scale $\mu_8$, we have found that there are only $13$ such models ($(n_\mathrm{X} , n_H)$ sets). Of those, $12$ work for the case where $\mu_{_\mathrm{VLF}} = 1~\mathrm{TeV}$, while only $9$ do so for the case where $\mu_{_\mathrm{VLF}} = 10~\mathrm{TeV}$. Our results are presented in \cref{tab:1TeV,tab:10TeV}. 
\begin{table}[htb]
	\begin{center}
		\begin{tabular}{c|cccccccc}
			\toprule                     
			Model & $\log_{10}\tfrac{\mu_8}{\mathrm{GeV}}$ & $\log_{10}\tfrac{\mu_{_\mathrm{PS}}}{\mathrm{GeV}}$ & $\alpha_8^{-1}\left(\mu_8\right)$ & $n_\mathrm{H}$ & $n_\mathrm{U_V}$ & $n_\mathrm{D_V}$ & $n_\mathrm{L_V}$ & $n_\mathrm{E_V}$ \\  
			\midrule
			$1$ & $19.05$ & $19.01$ & $47.51$ & $0$ & $1$ & $0$ & $0$ & $0$ \\
			$2$ & $18.79$ & $18.50$ & $42.77$ & $0$ & $1$ & $0$ & $0$ & $1$ \\
			$3$ & $18.87$ & $18.66$ & $44.32$ & $0$ & $0$ & $1$ & $0$ & $1$ \\
			$4$ & $18.54$ & $18.02$ & $38.31$ & $0$ & $1$ & $0$ & $0$ & $2$ \\
			$5$ & $18.26$ & $18.24$ & $46.04$ & $1$ & $1$ & $0$ & $0$ & $2$ \\
			$6$ & $18.62$ & $18.17$ & $39.77$ & $0$ & $0$ & $1$ & $0$ & $2$ \\
			$7$ & $18.30$ & $17.57$ & $34.13$ & $0$ & $1$ & $0$ & $0$ & $3$ \\
			$8$ & $18.03$ & $17.77$ & $41.57$ & $1$ & $1$ & $0$ & $0$ & $3$ \\
			$9$ & $18.38$ & $17.71$ & $35.49$ & $0$ & $0$ & $1$ & $0$ & $3$ \\
			$10$ & $18.11$ & $17.93$ & $43.03$ & $1$ & $0$ & $1$ & $0$ & $3$ \\
			$11$ & $18.96$ & $18.82$ & $41.90$ & $0$ & $2$ & $0$ & $1$ & $3$ \\
			$12$ & $19.05$ & $18.99$ & $43.47$ & $0$ & $1$ & $1$ & $1$ & $3$ \\
			\bottomrule
		\end{tabular} 
		\caption{Benchmark models consistent with the unification of gauge couplings at $\mu_8$ for $\mu_{_\mathrm{VLF}} = 1~\mathrm{TeV}$. All viable scenarios have $n_\mathrm{Q_V} = 0$.}
		\label{tab:1TeV}  
	\end{center}
\end{table}
\begin{table}[htb]
	\begin{center}
		\begin{tabular}{c|cccccccc}
			\toprule                     
			Model & $\log_{10}\tfrac{\mu_8}{\mathrm{GeV}}$ & $\log_{10}\tfrac{\mu_{_\mathrm{PS}}}{\mathrm{GeV}}$ & $\alpha_8^{-1}\left(\mu_8\right)$ & $n_\mathrm{H}$ & $n_\mathrm{U_V}$ & $n_\mathrm{D_V}$ & $n_\mathrm{L_V}$ & $n_\mathrm{E_V}$ \\
			\midrule
		$2$ & $18.69$ & $18.50$ & $45.89$ & $0$ & $1$ & $0$ & $0$ & $1$ \\
		$3$ & $18.77$ & $18.66$ & $45.89$ & $0$ & $0$ & $1$ & $0$ & $1$ \\
		$4$ & $18.46$ & $18.05$ & $40.27$ & $0$ & $1$ & $0$ & $0$ & $2$ \\
		$6$ & $18.54$ & $18.20$ & $41.63$ & $0$ & $0$ & $1$ & $0$ & $2$ \\
		$7$ & $18.24$ & $17.63$ & $36.36$ & $0$ & $1$ & $0$ & $0$ & $3$ \\
		$8$ & $17.97$ & $17.84$ & $43.83$ & $1$ & $1$ & $0$ & $0$ & $3$ \\
		$9$ & $18.31$ & $17.77$ & $37.63$ & $0$ & $0$ & $1$ & $0$ & $3$ \\
		$10$ & $18.04$ & $17.98$ & $45.20$ & $1$ & $0$ & $1$ & $0$ & $3$ \\
		$13$ & $18.85$ & $18.80$ & $43.63$ & $0$ & $1$ & $0$ & $1$ & $3$ \\
			\bottomrule
		\end{tabular}
		\caption{Benchmark models consistent with the unification of gauge couplings at $\mu_8$ for $\mu_{_\mathrm{VLF}} = 10~\mathrm{TeV}$. All viable scenarios have $n_\mathrm{Q_V} = 0$.}
		\label{tab:10TeV}  
	\end{center}
\end{table}

\begin{figure}[]
    \subfloat[Model 2: $n_\mathrm{U_V} = n_\mathrm{E_V} = 1$.]{{\hspace{-1cm}\includegraphics[width=0.55\textwidth]{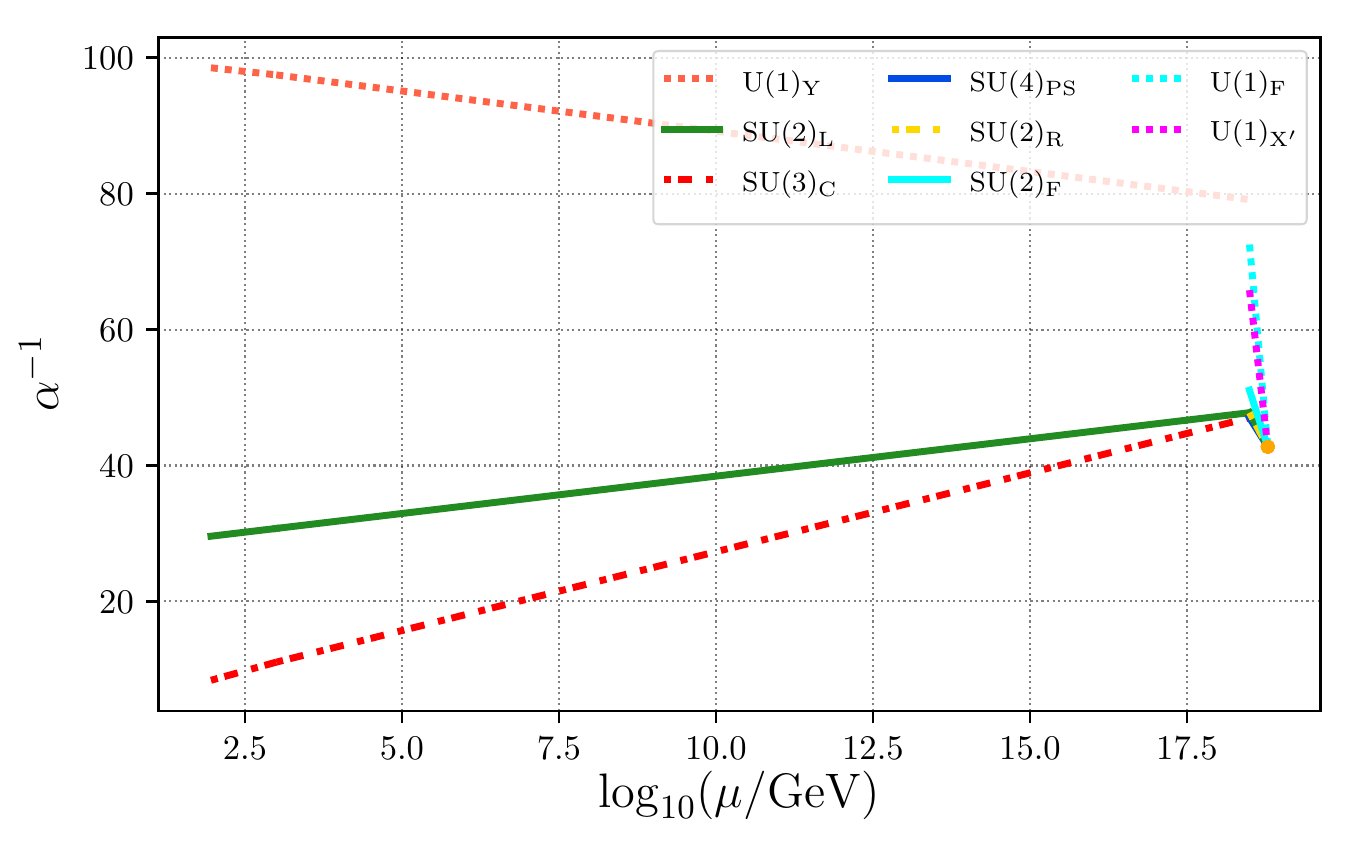} }} 
	\subfloat[Model 3: $n_\mathrm{D_V} = n_\mathrm{E_V} = 1$.]{{\includegraphics[width=0.55\textwidth]{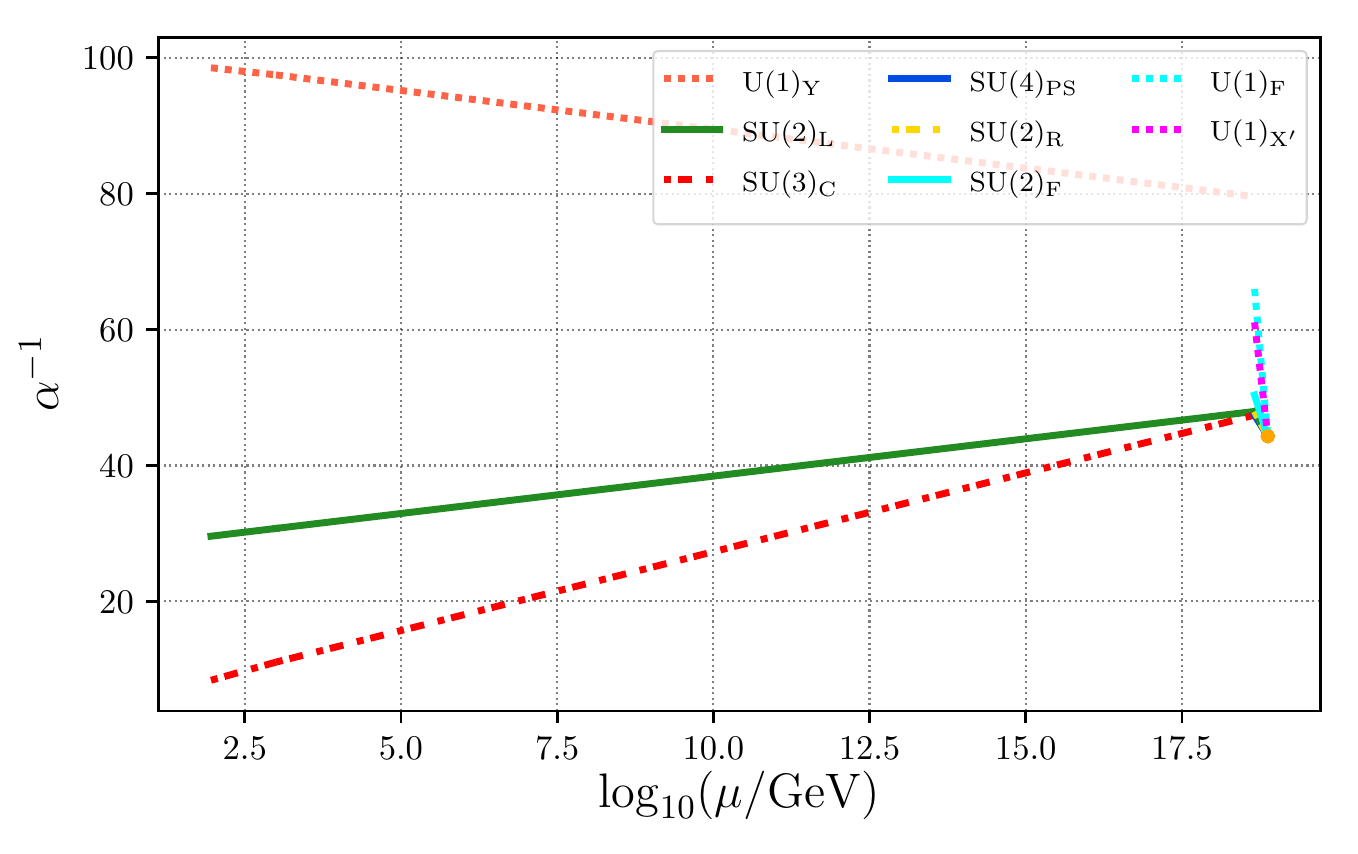} }}\\
	\subfloat[Model 7: $n_\mathrm{U_V} = 1,~ n_\mathrm{E_V} = 3$.]{{\hspace*{-1.0cm}\includegraphics[width=0.55\textwidth]{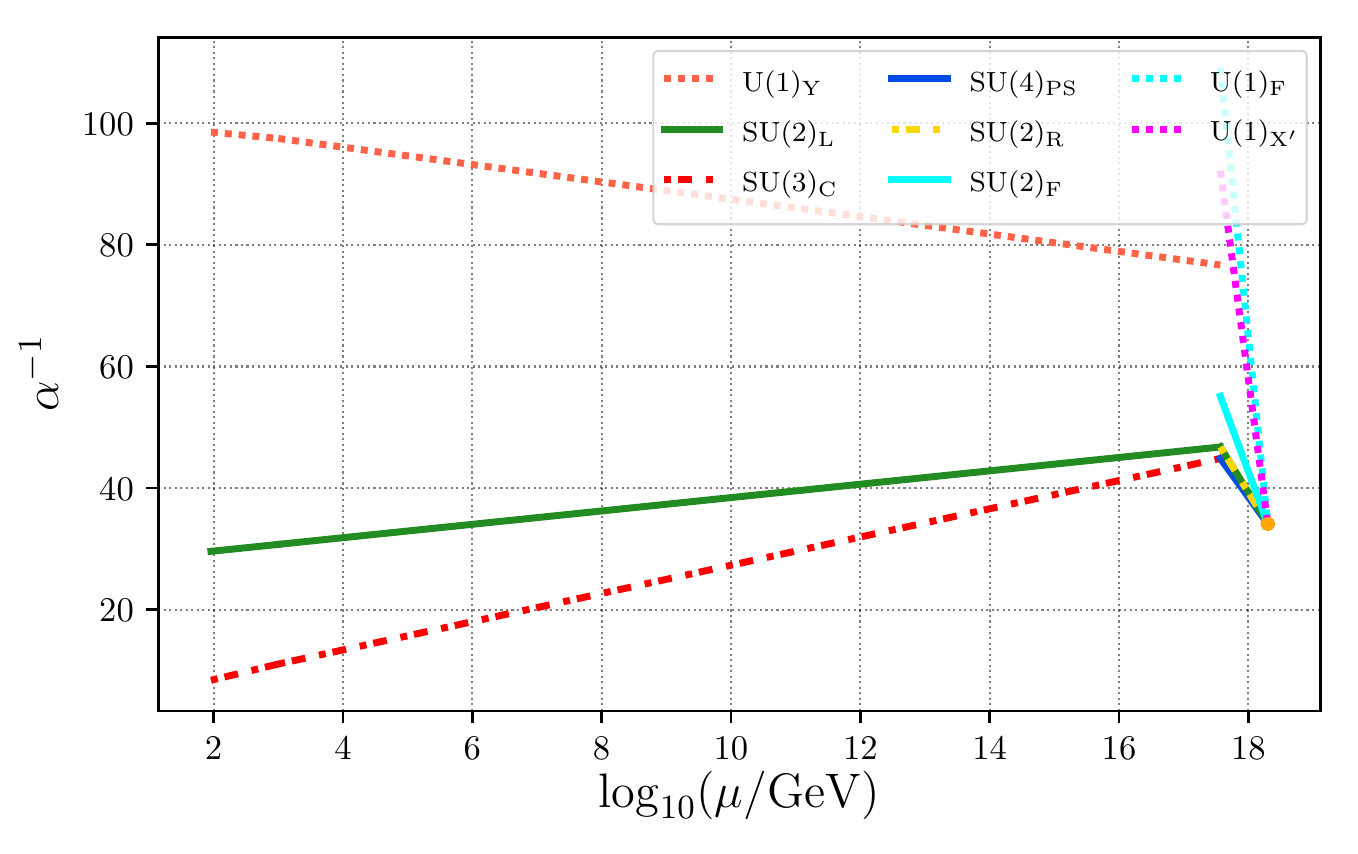} }} 
	\subfloat[Model 9: $n_\mathrm{D_V} = 1,~ n_\mathrm{E_V} = 3$.]{{\includegraphics[width=0.55\textwidth]{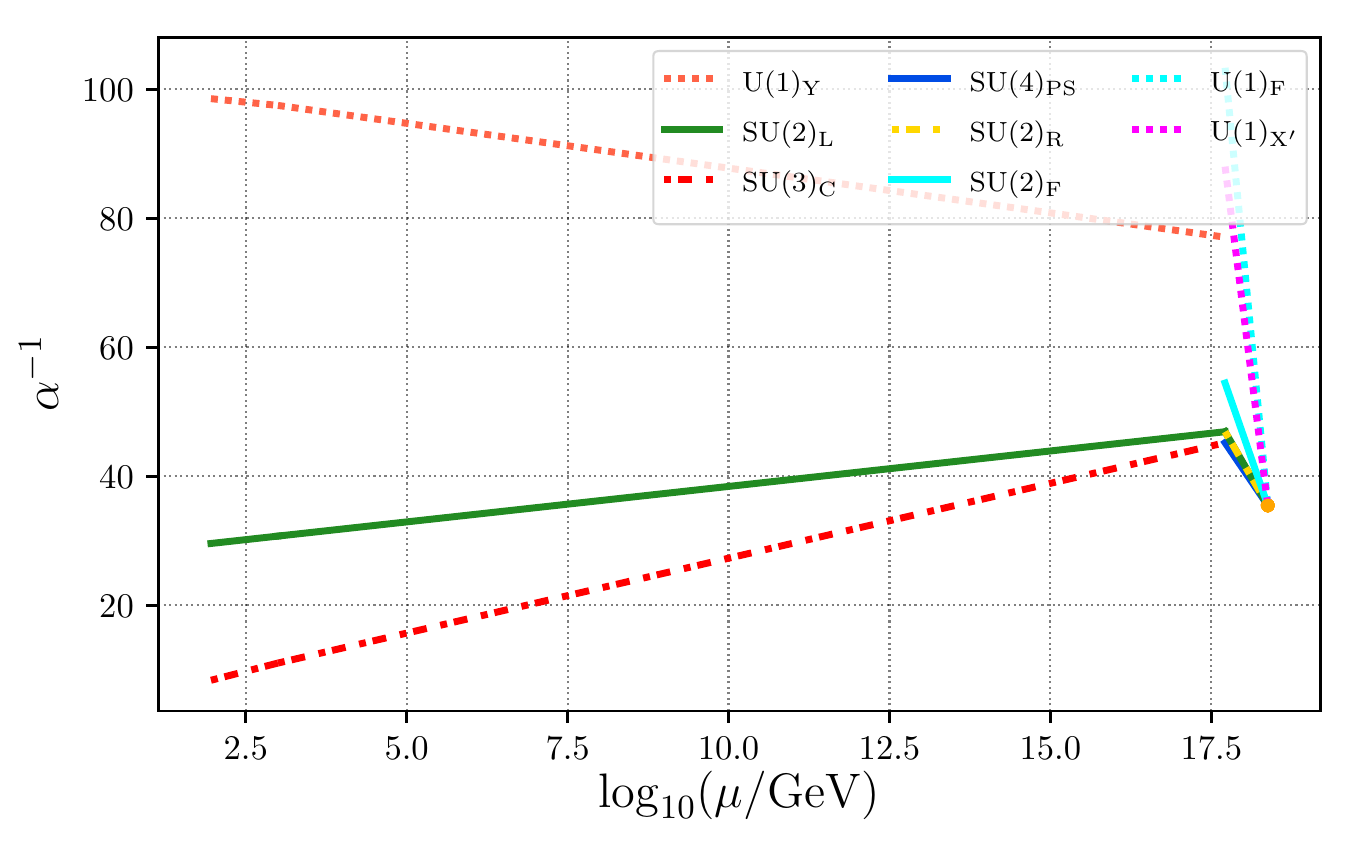} }}\\
	\subfloat[Model 10: $n_\mathrm{U_V} = 1,~ n_\mathrm{E_V} = 3,~n_\mathrm{H} = 1$.]{{\hspace*{-1.0cm}\includegraphics[width=0.55\textwidth]{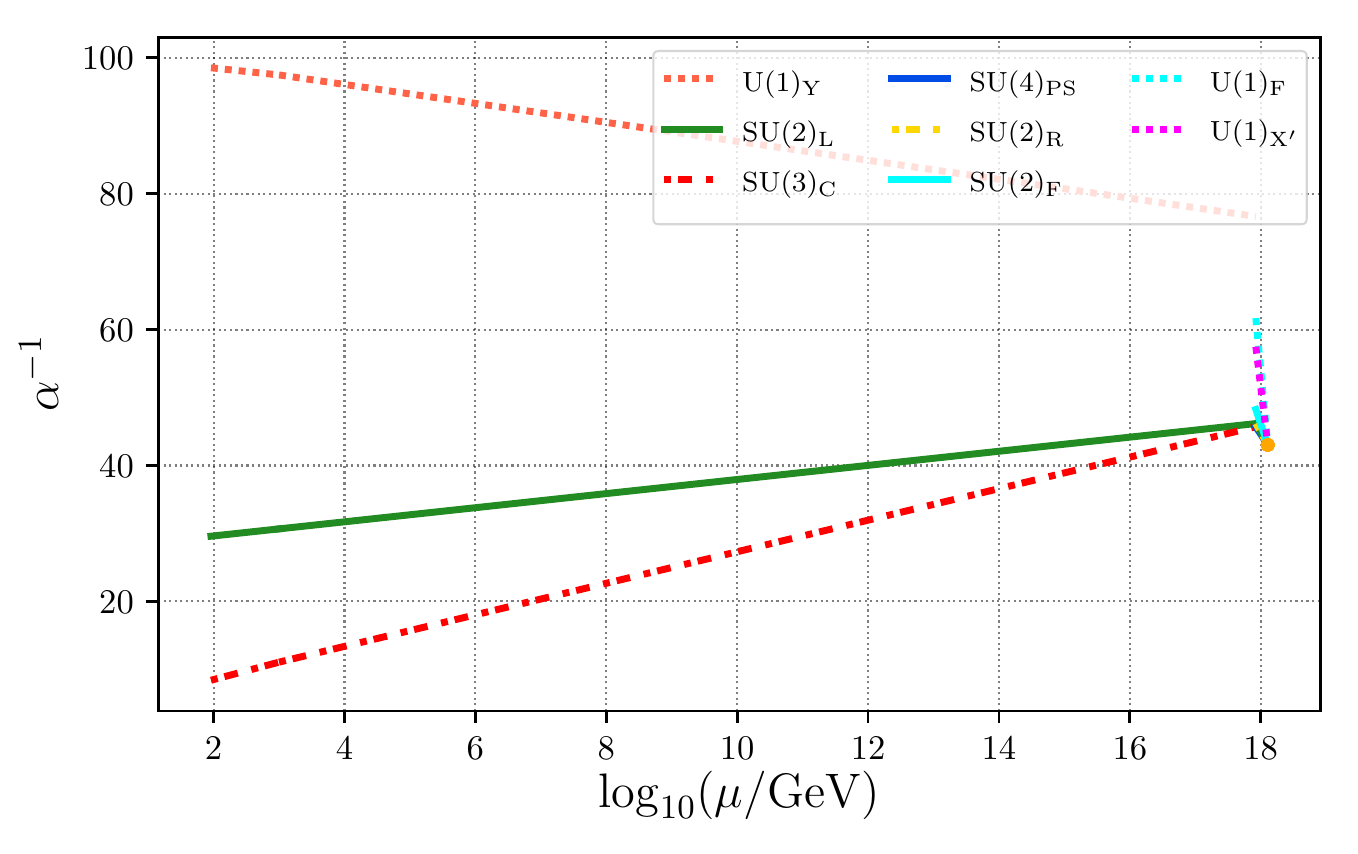} }} 
	\subfloat[Model 11: $n_\mathrm{U_V} = 2,~ n_\mathrm{E_V} = 3,~n_\mathrm{L_V} = 1$.]{{\includegraphics[width=0.55\textwidth]{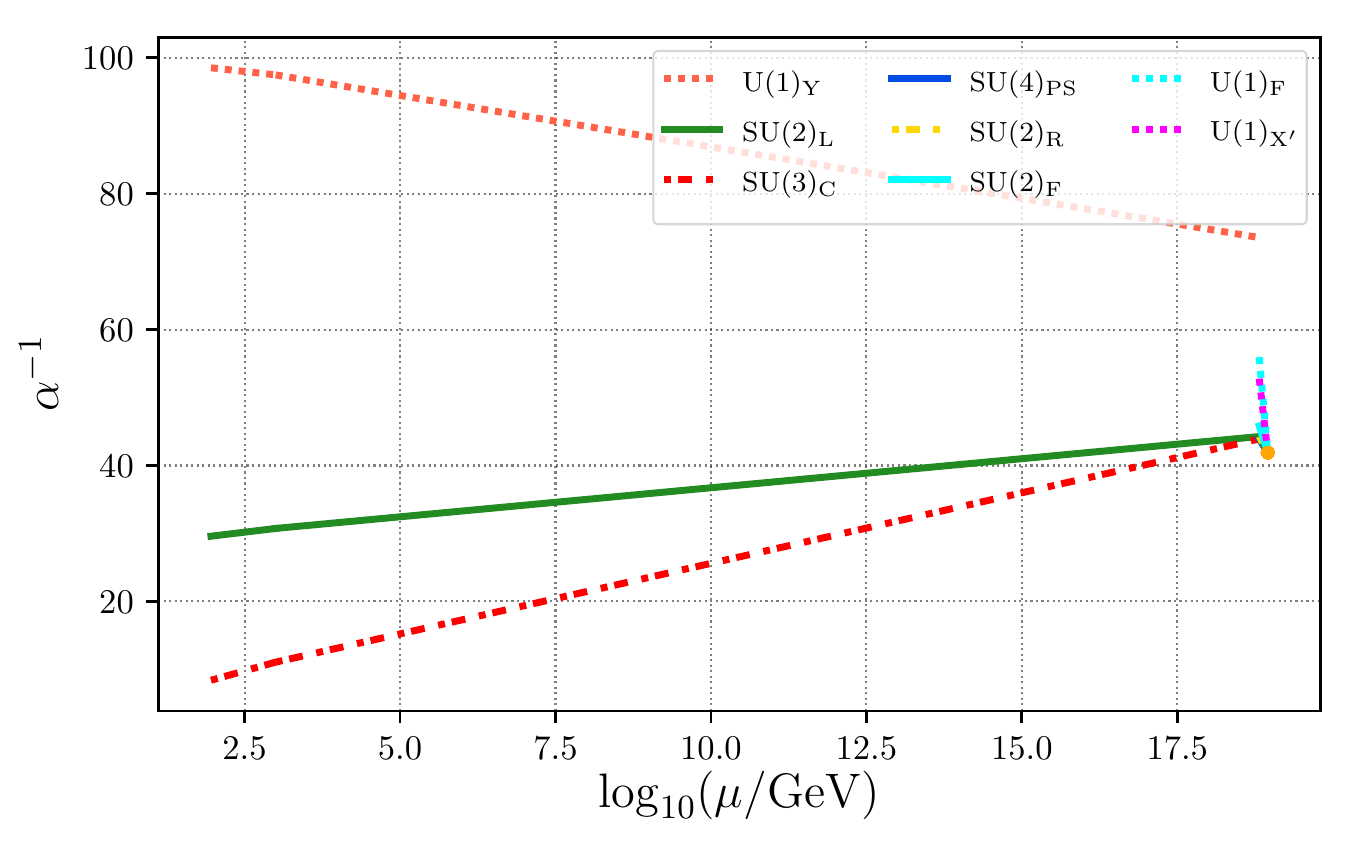} }} 
	\caption{RG evolution of the gauge couplings in the non-minimal Pati-Salam GUT for six viable low-scale models compatible with exact unification of all interactions. In all cases shown above we have $\mu_{_\mathrm{VLF}} = 1~\mathrm{TeV}$ as in \cref{tab:1TeV}. The orange dot represents the unification point with a universal gauge coupling $g_8$.}
	\label{fig:RGEs}
\end{figure}

We have found that at least one generation of $\SU{2}{L}$ singlet VLQs is needed (Model 1) while none of the viable scenarios allow for $\SU{2}{L}$ doublet VLQs in the low-energy scale spectrum. Furthermore, Model 1 is the only one not containing VLLs while only Models 11, 12 and 13 allow for $\SU{2}{L}$ doublet VLLs. Model 11 is also the only one with more than one generation of $\mathrm{U_V}$ VLQs. The compactification scale always satisfies $\mu_8 > 10^{18}~\mathrm{GeV}$ while the Wilson line scale $\mu_{_\mathrm{PS}} > 10^{17.5}~\mathrm{GeV}$, not too far from $\mu_8$. The value of the universal $\mathrm{E}_8$ gauge coupling is also limited to a small range $36 \lesssim \alpha_8^{-1}\left(\mu_8\right) \lesssim 48$. Finally, we show in \cref{fig:RGEs} six representative examples of models 2, 3, 7, 9, 10 and 11, always for the case of $\mu_{_\mathrm{VLF}} = 1~\mathrm{TeV}$.\\

\newpage

\section{Conclusions}
\label{sec:con}

We have studied an anomaly-free realization of a non-minimal Pati-Salam GUT unified under the $\mathrm{E}_8$ symmetry. The 10-dimensional theory contains one vector $\bm{248}$ as well as five chiral $\bm{248}$-plets. The 4-dimensional limit of the $\mathrm{E}_8$ theory upon $\mathbb{Z}_6 \times \mathbb{Z}_2$ orbifold compactification contains massless zero modes that can describe all SM gauge fields, chiral matter and right-handed neutrinos. We have shown that, in the limit of approximately degenerate superpotential mass parameters, our Pati-Salam GUT can naturally contain vector-like fermions at low-energy scales and at the reach of LHC or future collider experiments. In particular, the unification of the gauge couplings under $\mathrm{E}_8$ requires that such exotic fermions can either be $\SU{2}{L}$ singlet VLQs or both singlet and doublet VLLs. In total, we have found 13 viable models with New Physics manifest at $1~\mathrm{TeV}$ or at $10~\mathrm{TeV}$, making our model falsifiable at the LHC or future colliders.

\section*{Acknowledgments}

The authors want to thank Stephen F. King for thorough and insightful discussions about the problems addressed in this manuscript.
A.A. acknowledges support form CONACYT-SNI (M\'exico).
A.P.M.~is supported by the Center for Research and Development in Mathematics and Applications (CIDMA) through the Portuguese Foundation for Science and Technology (FCT - Fundação para a Ciência e a Tecnologia), references UIDB/04106/2020 and UIDP/04106/2020. A.P.M.~is also supported by the projects PTDC/FIS-PAR/31000/2017, CERN/FIS-PAR/0027/2019, CERN/FISPAR/0002/2017 and by national funds (OE), through FCT, I.P., in the scope of the framework contract foreseen in the numbers 4, 5 and 6 of the article 23, of the Decree-Law 57/2016, of August 29, changed by Law 57/2017, of July 19. 
R.P.~is supported in part by the Swedish Research Council grants, contract numbers 621-2013-4287 and 2016-05996, as well as by the European Research Council (ERC) under the European Union's Horizon 2020 research and innovation programme (grant agreement No 668679).

\bibliographystyle{JHEP}
\bibliography{bib}

\providecommand{\href}[2]{#2}\begingroup\raggedright\begin{thebibliography}{10}

\bibitem{Georgi:1974sy}
H.~Georgi and S.~L. Glashow, \emph{{Unity of All Elementary Particle Forces}},
  \href{https://doi.org/10.1103/PhysRevLett.32.438}{\emph{Phys. Rev. Lett.}
  {\bfseries 32} (1974) 438}.

\bibitem{Fritzsch:1974nn}
H.~Fritzsch and P.~Minkowski, \emph{{Unified Interactions of Leptons and
  Hadrons}}, \href{https://doi.org/10.1016/0003-4916(75)90211-0}{\emph{Annals
  Phys.} {\bfseries 93} (1975) 193}.

\bibitem{King:2005my}
S.~F. King, S.~Moretti and R.~Nevzorov, \emph{{Exceptional supersymmetric
  standard model}},
  \href{https://doi.org/10.1016/j.physletb.2005.12.070}{\emph{Phys. Lett.}
  {\bfseries B634} (2006) 278}
  [\href{https://arxiv.org/abs/hep-ph/0511256}{{\ttfamily hep-ph/0511256}}].

\bibitem{Buchmuller:1985rc}
W.~Buchmuller and O.~Napoly, \emph{{Exceptional Coset Spaces and the Spectrum
  of Quarks and Leptons}},
  \href{https://doi.org/10.1016/0370-2693(85)90212-6}{\emph{Phys. Lett. B}
  {\bfseries 163} (1985) 161}.

\bibitem{Koca:1982zi}
M.~Koca, \emph{{EXPLICIT REALIZATION OF E8.}},
  \href{https://doi.org/10.1007/3-540-12291-5_55}{\emph{Lect. Notes Phys.}
  {\bfseries 180} (2005) 356}.

\bibitem{Slansky:1981yr}
R.~Slansky, \emph{{Group Theory for Unified Model Building}},
  \href{https://doi.org/10.1016/0370-1573(81)90092-2}{\emph{Phys. Rept.}
  {\bfseries 79} (1981) 1}.

\bibitem{King:2001uz}
S.~King and G.~G. Ross, \emph{{Fermion masses and mixing angles from SU(3)
  family symmetry}},
  \href{https://doi.org/10.1016/S0370-2693(01)01139-X}{\emph{Phys. Lett. B}
  {\bfseries 520} (2001) 243}
  [\href{https://arxiv.org/abs/hep-ph/0108112}{{\ttfamily hep-ph/0108112}}].

\bibitem{King:2017guk}
S.~King, \emph{{Unified Models of Neutrinos, Flavour and CP Violation}},
  \href{https://doi.org/10.1016/j.ppnp.2017.01.003}{\emph{Prog. Part. Nucl.
  Phys.} {\bfseries 94} (2017) 217}
  [\href{https://arxiv.org/abs/1701.04413}{{\ttfamily 1701.04413}}].

\bibitem{Hagedorn:2010th}
C.~Hagedorn, S.~F. King and C.~Luhn, \emph{{A SUSY GUT of Flavour with $S_4
  \times SU(5)$ to NLO}},
  \href{https://doi.org/10.1007/JHEP06(2010)048}{\emph{JHEP} {\bfseries 06}
  (2010) 048} [\href{https://arxiv.org/abs/1003.4249}{{\ttfamily 1003.4249}}].

\bibitem{Antusch:2014poa}
S.~Antusch, I.~de~Medeiros~Varzielas, V.~Maurer, C.~Sluka and M.~Spinrath,
  \emph{{Towards predictive flavour models in SUSY SU(5) GUTs with
  doublet-triplet splitting}},
  \href{https://doi.org/10.1007/JHEP09(2014)141}{\emph{JHEP} {\bfseries 09}
  (2014) 141} [\href{https://arxiv.org/abs/1405.6962}{{\ttfamily 1405.6962}}].

\bibitem{Bjorkeroth:2015ora}
F.~Bj\"orkeroth, F.~J. de~Anda, I.~de~Medeiros~Varzielas and S.~F. King,
  \emph{{Towards a complete A$_{4} \times$ SU(5) SUSY GUT}},
  \href{https://doi.org/10.1007/JHEP06(2015)141}{\emph{JHEP} {\bfseries 06}
  (2015) 141} [\href{https://arxiv.org/abs/1503.03306}{{\ttfamily
  1503.03306}}].

\bibitem{Bjorkeroth:2015uou}
F.~Bj\"orkeroth, F.~J. de~Anda, I.~de~Medeiros~Varzielas and S.~F. King,
  \emph{{Towards a complete $\Delta(27) \times SO(10)$ SUSY GUT}},
  \href{https://doi.org/10.1103/PhysRevD.94.016006}{\emph{Phys. Rev. D}
  {\bfseries 94} (2016) 016006}
  [\href{https://arxiv.org/abs/1512.00850}{{\ttfamily 1512.00850}}].

\bibitem{Bjorkeroth:2017ybg}
F.~Bj\"orkeroth, F.~J. de~Anda, S.~F. King and E.~Perdomo, \emph{{A natural
  $S_{4} \times SO(10)$ model of flavour}},
  \href{https://doi.org/10.1007/JHEP10(2017)148}{\emph{JHEP} {\bfseries 10}
  (2017) 148} [\href{https://arxiv.org/abs/1705.01555}{{\ttfamily
  1705.01555}}].

\bibitem{deAnda:2017yeb}
F.~J. de~Anda, S.~F. King and E.~Perdomo, \emph{{$\mathbf{SO(10)}\times
  \mathbf{S_4}$ grand unified theory of flavour and leptogenesis}},
  \href{https://doi.org/10.1007/JHEP12(2017)075}{\emph{JHEP} {\bfseries 12}
  (2017) 075} [\href{https://arxiv.org/abs/1710.03229}{{\ttfamily
  1710.03229}}].

\bibitem{CarcamoHernandez:2020owa}
A.~C\'arcamo~Hern\'andez, D.~Huong, S.~Kovalenko, A.~P. Morais, R.~Pasechnik
  and I.~Schmidt, \emph{{How low-scale trinification sheds light in the flavor
  hierarchies, neutrino puzzle, dark matter, and leptogenesis}},
  \href{https://doi.org/10.1103/PhysRevD.102.095003}{\emph{Phys. Rev. D}
  {\bfseries 102} (2020) 095003}
  [\href{https://arxiv.org/abs/2004.11450}{{\ttfamily 2004.11450}}].

\bibitem{Morais:2020ypd}
A.~P. Morais, R.~Pasechnik and W.~Porod, \emph{{Prospects for New Physics from
  gauge Left-Right-Colour-Family Grand Unification}},
  \href{https://arxiv.org/abs/2001.06383}{{\ttfamily 2001.06383}}.

\bibitem{Morais:2020odg}
A.~P. Morais, R.~Pasechnik and W.~Porod, \emph{{Grand Unified origin of gauge
  interactions and families replication in the Standard Model}},
  \href{https://arxiv.org/abs/2001.04804}{{\ttfamily 2001.04804}}.

\bibitem{Camargo-Molina:2016yqm}
J.~E. Camargo-Molina, A.~P. Morais, A.~Ordell, R.~Pasechnik, M.~O. Sampaio and
  J.~Wess{\'e}n, \emph{{Reviving trinification models through an E6 -extended
  supersymmetric GUT}},
  \href{https://doi.org/10.1103/PhysRevD.95.075031}{\emph{Phys. Rev.}
  {\bfseries D95} (2017) 075031}
  [\href{https://arxiv.org/abs/1610.03642}{{\ttfamily 1610.03642}}].

\bibitem{Camargo-Molina:2017kxd}
J.~E. Camargo-Molina, A.~P. Morais, A.~Ordell, R.~Pasechnik and J.~Wessén,
  \emph{{Scale hierarchies, symmetry breaking and particle spectra in
  SU(3)-family extended SUSY trinification}},
  \href{https://doi.org/10.1103/PhysRevD.99.035041}{\emph{Phys. Rev.}
  {\bfseries D99} (2019) 035041}
  [\href{https://arxiv.org/abs/1711.05199}{{\ttfamily 1711.05199}}].

\bibitem{Camargo-Molina:2016bwm}
J.~E. Camargo-Molina, A.~P. Morais, R.~Pasechnik and J.~Wess{\'e}n, \emph{{On a
  radiative origin of the Standard Model from Trinification}},
  \href{https://doi.org/10.1007/JHEP09(2016)129}{\emph{JHEP} {\bfseries 09}
  (2016) 129} [\href{https://arxiv.org/abs/1606.03492}{{\ttfamily
  1606.03492}}].

\bibitem{Altarelli:2008bg}
G.~Altarelli, F.~Feruglio and C.~Hagedorn, \emph{{A SUSY SU(5) Grand Unified
  Model of Tri-Bimaximal Mixing from A$_4$}},
  \href{https://doi.org/10.1088/1126-6708/2008/03/052}{\emph{JHEP} {\bfseries
  03} (2008) 052} [\href{https://arxiv.org/abs/0802.0090}{{\ttfamily
  0802.0090}}].

\bibitem{Burrows:2009pi}
T.~Burrows and S.~King, \emph{{A(4) Family Symmetry from SU(5) SUSY GUTs in
  6d}}, \href{https://doi.org/10.1016/j.nuclphysb.2010.04.002}{\emph{Nucl.
  Phys. B} {\bfseries 835} (2010) 174}
  [\href{https://arxiv.org/abs/0909.1433}{{\ttfamily 0909.1433}}].

\bibitem{Burrows:2010wz}
T.~Burrows and S.~King, \emph{{$A_4$ x SU(5) SUSY GUT of Flavour in 8d}},
  \href{https://doi.org/10.1016/j.nuclphysb.2010.08.018}{\emph{Nucl. Phys. B}
  {\bfseries 842} (2011) 107}
  [\href{https://arxiv.org/abs/1007.2310}{{\ttfamily 1007.2310}}].

\bibitem{deAnda:2018oik}
F.~J. de~Anda and S.~F. King, \emph{{An $S_4 \times SU(5)$ SUSY GUT of flavour
  in 6d}}, \href{https://doi.org/10.1007/JHEP07(2018)057}{\emph{JHEP}
  {\bfseries 07} (2018) 057}
  [\href{https://arxiv.org/abs/1803.04978}{{\ttfamily 1803.04978}}].

\bibitem{Altarelli:2006kg}
G.~Altarelli, F.~Feruglio and Y.~Lin, \emph{{Tri-bimaximal neutrino mixing from
  orbifolding}},
  \href{https://doi.org/10.1016/j.nuclphysb.2007.03.042}{\emph{Nucl. Phys. B}
  {\bfseries 775} (2007) 31}
  [\href{https://arxiv.org/abs/hep-ph/0610165}{{\ttfamily hep-ph/0610165}}].

\bibitem{Adulpravitchai:2010na}
A.~Adulpravitchai and M.~A. Schmidt, \emph{{Flavored Orbifold GUT - an SO(10) x
  S4 model}}, \href{https://doi.org/10.1007/JHEP01(2011)106}{\emph{JHEP}
  {\bfseries 01} (2011) 106} [\href{https://arxiv.org/abs/1001.3172}{{\ttfamily
  1001.3172}}].

\bibitem{Adulpravitchai:2009id}
A.~Adulpravitchai, A.~Blum and M.~Lindner, \emph{{Non-Abelian Discrete Flavor
  Symmetries from T**2/Z(N) Orbifolds}},
  \href{https://doi.org/10.1088/1126-6708/2009/07/053}{\emph{JHEP} {\bfseries
  07} (2009) 053} [\href{https://arxiv.org/abs/0906.0468}{{\ttfamily
  0906.0468}}].

\bibitem{Asaka:2001eh}
T.~Asaka, W.~Buchmuller and L.~Covi, \emph{{Gauge unification in
  six-dimensions}},
  \href{https://doi.org/10.1016/S0370-2693(01)01324-7}{\emph{Phys. Lett. B}
  {\bfseries 523} (2001) 199}
  [\href{https://arxiv.org/abs/hep-ph/0108021}{{\ttfamily hep-ph/0108021}}].

\bibitem{deAnda:2019jxw}
F.~J. de~Anda, J.~W. Valle and C.~A. Vaquera-Araujo, \emph{{Flavour and CP
  predictions from orbifold compactification}},
  \href{https://doi.org/10.1016/j.physletb.2019.135195}{\emph{Phys. Lett. B}
  {\bfseries 801} (2020) 135195}
  [\href{https://arxiv.org/abs/1910.05605}{{\ttfamily 1910.05605}}].

\bibitem{deAnda:2018ecu}
F.~J. de~Anda, S.~F. King and E.~Perdomo, \emph{{$SU(5)$ grand unified theory
  with $A_4$ modular symmetry}},
  \href{https://doi.org/10.1103/PhysRevD.101.015028}{\emph{Phys. Rev. D}
  {\bfseries 101} (2020) 015028}
  [\href{https://arxiv.org/abs/1812.05620}{{\ttfamily 1812.05620}}].

\bibitem{deAnda:2018yfp}
F.~J. de~Anda and S.~F. King, \emph{{$SU(3) \times SO(10)$ in 6d}},
  \href{https://doi.org/10.1007/JHEP10(2018)128}{\emph{JHEP} {\bfseries 10}
  (2018) 128} [\href{https://arxiv.org/abs/1807.07078}{{\ttfamily
  1807.07078}}].

\bibitem{Ibanez:1987pj}
L.~E. Ibanez, J.~Mas, H.-P. Nilles and F.~Quevedo, \emph{{Heterotic Strings in
  Symmetric and Asymmetric Orbifold Backgrounds}},
  \href{https://doi.org/10.1016/0550-3213(88)90166-6}{\emph{Nucl. Phys. B}
  {\bfseries 301} (1988) 157}.

\bibitem{Parr:2020oar}
E.~Parr, P.~K. Vaudrevange and M.~Wimmer, \emph{{Predicting the orbifold origin
  of the MSSM}}, \href{https://doi.org/10.1002/prop.202000032}{\emph{Fortsch.
  Phys.} {\bfseries 68} (2020) 2000032}
  [\href{https://arxiv.org/abs/2003.01732}{{\ttfamily 2003.01732}}].

\bibitem{Adler:2002yg}
S.~L. Adler, \emph{{Should E(8) SUSY Yang-Mills be reconsidered as a family
  unification model?}},
  \href{https://doi.org/10.1016/S0370-2693(02)01596-4}{\emph{Phys. Lett.}
  {\bfseries B533} (2002) 121}
  [\href{https://arxiv.org/abs/hep-ph/0201009}{{\ttfamily hep-ph/0201009}}].

\bibitem{Adler:2004uj}
S.~L. Adler, \emph{{Further thoughts on supersymmetric E(8) as a family and
  grand unification theory}},
  \href{https://arxiv.org/abs/hep-ph/0401212}{{\ttfamily hep-ph/0401212}}.

\bibitem{Garibaldi:2016zgm}
S.~Garibaldi, \emph{{$\mathrm{E}_8$, the most exceptional group}},
  \href{https://doi.org/10.1090/bull/1540}{\emph{Bull. Am. Math. Soc.}
  {\bfseries 53} (2016) 643}
  [\href{https://arxiv.org/abs/1605.01721}{{\ttfamily 1605.01721}}].

\bibitem{Thomas:1985be}
S.~Thomas, \emph{{SOFTLY BROKEN N=4 AND E8}},
  \href{https://doi.org/10.1088/0305-4470/19/7/016}{\emph{J. Phys. A}
  {\bfseries 19} (1986) 1141}.

\bibitem{Konshtein:1980km}
S.~Konshtein and E.~Fradkin, \emph{{ASYMPTOTICALLY SUPERSYMMETRIC MODEL OF
  UNIFIED INTERACTION BASED ON E8. (IN RUSSIAN)}}, {\emph{Pisma Zh. Eksp. Teor.
  Fiz.} {\bfseries 32} (1980) 575}.

\bibitem{Baaklini:1980fv}
N.~Baaklini, \emph{{SUPERSYMMETRIC EXCEPTIONAL GAUGE UNIFICATION}},
  \href{https://doi.org/10.1103/PhysRevD.22.3118}{\emph{Phys. Rev. D}
  {\bfseries 22} (1980) 3118}.

\bibitem{Baaklini:1980uq}
N.~Baaklini, \emph{{SUPERGRAND UNIFICATION IN E8}},
  \href{https://doi.org/10.1016/0370-2693(80)90999-5}{\emph{Phys. Lett. B}
  {\bfseries 91} (1980) 376}.

\bibitem{Barr:1987pu}
S.~M. Barr, \emph{{$E_8$ family unification, mirror fermions, and new
  low-energy physics}},
  \href{https://doi.org/10.1103/PhysRevD.37.204}{\emph{Phys. Rev. D} {\bfseries
  37} (1988) 204}.

\bibitem{Bars:1980mb}
I.~Bars and M.~Gunaydin, \emph{{Grand Unification With the Exceptional Group
  E8}}, \href{https://doi.org/10.1103/PhysRevLett.45.859}{\emph{Phys. Rev.
  Lett.} {\bfseries 45} (1980) 859}.

\bibitem{Koca:1981xd}
M.~Koca, \emph{{ON TUMBLING E8}},
  \href{https://doi.org/10.1016/0370-2693(81)91150-3}{\emph{Phys. Lett. B}
  {\bfseries 107} (1981) 73}.

\bibitem{Mahapatra:1988gc}
S.~Mahapatra and B.~Deo, \emph{{SUPERGRAVITY INDUCED E(8) GAUGE HIERARCHIES}},
  \href{https://doi.org/10.1103/PhysRevD.38.3554}{\emph{Phys. Rev. D}
  {\bfseries 38} (1988) 3554}.

\bibitem{Ong:1984ej}
C.-L. Ong, \emph{{Supersymmetric Models for Quarks and Leptons With Nonlinearly
  Realized E8 Symmetry}},
  \href{https://doi.org/10.1103/PhysRevD.31.3271}{\emph{Phys. Rev. D}
  {\bfseries 31} (1985) 3271}.

\bibitem{Olive:1982ai}
D.~I. Olive and P.~C. West, \emph{{The $N=4$ Supersymmetric $E(8)$ Gauge Theory
  and Coset Space: Dimensional Reduction}},
  \href{https://doi.org/10.1016/0550-3213(83)90086-X}{\emph{Nucl. Phys. B}
  {\bfseries 217} (1983) 248}.

\bibitem{ArkaniHamed:2001tb}
N.~Arkani-Hamed, T.~Gregoire and J.~G. Wacker, \emph{{Higher dimensional
  supersymmetry in 4-D superspace}},
  \href{https://doi.org/10.1088/1126-6708/2002/03/055}{\emph{JHEP} {\bfseries
  03} (2002) 055} [\href{https://arxiv.org/abs/hep-th/0101233}{{\ttfamily
  hep-th/0101233}}].

\bibitem{Brink:1976bc}
L.~Brink, J.~H. Schwarz and J.~Scherk, \emph{{Supersymmetric Yang-Mills
  Theories}}, \href{https://doi.org/10.1016/0550-3213(77)90328-5}{\emph{Nucl.
  Phys. B} {\bfseries 121} (1977) 77}.

\bibitem{Aranda:2020noz}
A.~Aranda, F.~J. de~Anda and S.~F. King, \emph{{Exceptional Unification of
  Families and Forces}},
  \href{https://doi.org/10.1016/j.nuclphysb.2020.115209}{\emph{Nucl. Phys. B}
  {\bfseries 960} (2020) 115209}
  [\href{https://arxiv.org/abs/2005.03048}{{\ttfamily 2005.03048}}].

\bibitem{Aranda:2020zms}
A.~Aranda and F.~J. de~Anda, \emph{{Complete $E_8$ Unification in 10
  Dimensions}},  \href{https://arxiv.org/abs/2007.13248}{{\ttfamily
  2007.13248}}.

\bibitem{Pati:1973uk}
J.~C. Pati and A.~Salam, \emph{{Unified Lepton-Hadron Symmetry and a Gauge
  Theory of the Basic Interactions}},
  \href{https://doi.org/10.1103/PhysRevD.8.1240}{\emph{Phys. Rev. D} {\bfseries
  8} (1973) 1240}.

\bibitem{Dixon:1985jw}
L.~J. Dixon, J.~A. Harvey, C.~Vafa and E.~Witten, \emph{{Strings on
  Orbifolds}}, \href{https://doi.org/10.1016/0550-3213(85)90593-0}{\emph{Nucl.
  Phys.} {\bfseries B261} (1985) 678}.

\bibitem{Dixon:1986jc}
L.~J. Dixon, J.~A. Harvey, C.~Vafa and E.~Witten, \emph{{Strings on Orbifolds.
  2.}}, \href{https://doi.org/10.1016/0550-3213(86)90287-7}{\emph{Nucl. Phys.}
  {\bfseries B274} (1986) 285}.

\bibitem{GrootNibbelink:2017luf}
S.~Groot~Nibbelink, O.~Loukas, A.~M\"utter, E.~Parr and P.~K. Vaudrevange,
  \emph{{Tension Between a Vanishing Cosmological Constant and
  Non-Supersymmetric Heterotic Orbifolds}},
  \href{https://doi.org/10.1002/prop.202000044}{\emph{Fortsch. Phys.}
  {\bfseries 68} (2020) 2000044}
  [\href{https://arxiv.org/abs/1710.09237}{{\ttfamily 1710.09237}}].

\bibitem{deAnda:2019anb}
F.~J. De~Anda, S.~F. King, E.~Perdomo and P.~K. Vaudrevange, \emph{{Flavon
  alignments from orbifolding: $SU(5)\times SU(3)$ model with
  $\mathbb{T}^6/\Delta(54)$}},
  \href{https://doi.org/10.1007/JHEP12(2019)055}{\emph{JHEP} {\bfseries 12}
  (2019) 055} [\href{https://arxiv.org/abs/1910.04175}{{\ttfamily
  1910.04175}}].

\bibitem{Fischer:2012qj}
M.~Fischer, M.~Ratz, J.~Torrado and P.~K. Vaudrevange, \emph{{Classification of
  symmetric toroidal orbifolds}},
  \href{https://doi.org/10.1007/JHEP01(2013)084}{\emph{JHEP} {\bfseries 01}
  (2013) 084} [\href{https://arxiv.org/abs/1209.3906}{{\ttfamily 1209.3906}}].

\bibitem{Fischer:2013qza}
M.~Fischer, S.~Ramos-Sanchez and P.~K.~S. Vaudrevange, \emph{{Heterotic
  non-Abelian orbifolds}},
  \href{https://doi.org/10.1007/JHEP07(2013)080}{\emph{JHEP} {\bfseries 07}
  (2013) 080} [\href{https://arxiv.org/abs/1304.7742}{{\ttfamily 1304.7742}}].

\bibitem{Hosotani:1983xw}
Y.~Hosotani, \emph{{Dynamical Mass Generation by Compact Extra Dimensions}},
  \href{https://doi.org/10.1016/0370-2693(83)90170-3}{\emph{Phys. Lett. B}
  {\bfseries 126} (1983) 309}.

\bibitem{Hosotani:1983vn}
Y.~Hosotani, \emph{{Dynamical Gauge Symmetry Breaking as the Casimir Effect}},
  \href{https://doi.org/10.1016/0370-2693(83)90841-9}{\emph{Phys. Lett. B}
  {\bfseries 129} (1983) 193}.

\bibitem{Hosotani:2004wv}
Y.~Hosotani, S.~Noda and K.~Takenaga, \emph{{Dynamical gauge-Higgs unification
  in the electroweak theory}},
  \href{https://doi.org/10.1016/j.physletb.2004.12.029}{\emph{Phys. Lett. B}
  {\bfseries 607} (2005) 276}
  [\href{https://arxiv.org/abs/hep-ph/0410193}{{\ttfamily hep-ph/0410193}}].

\bibitem{Hosotani:2004ka}
Y.~Hosotani, S.~Noda and K.~Takenaga, \emph{{Dynamical gauge symmetry breaking
  and mass generation on the orbifold T**2 / Z(2)}},
  \href{https://doi.org/10.1103/PhysRevD.69.125014}{\emph{Phys. Rev. D}
  {\bfseries 69} (2004) 125014}
  [\href{https://arxiv.org/abs/hep-ph/0403106}{{\ttfamily hep-ph/0403106}}].

\bibitem{Haba:2004qf}
N.~Haba, Y.~Hosotani, Y.~Kawamura and T.~Yamashita, \emph{{Dynamical symmetry
  breaking in gauge Higgs unification on orbifold}},
  \href{https://doi.org/10.1103/PhysRevD.70.015010}{\emph{Phys. Rev. D}
  {\bfseries 70} (2004) 015010}
  [\href{https://arxiv.org/abs/hep-ph/0401183}{{\ttfamily hep-ph/0401183}}].

\bibitem{Haba:2002py}
N.~Haba, M.~Harada, Y.~Hosotani and Y.~Kawamura, \emph{{Dynamical rearrangement
  of gauge symmetry on the orbifold S1 / Z(2)}},
  \href{https://doi.org/10.1016/S0550-3213(03)00142-1}{\emph{Nucl. Phys. B}
  {\bfseries 657} (2003) 169}
  [\href{https://arxiv.org/abs/hep-ph/0212035}{{\ttfamily hep-ph/0212035}}].

\bibitem{Branco:1999fs}
G.~C. Branco, L.~Lavoura and J.~P. Silva, \emph{{CP Violation}}, {\emph{Int.
  Ser. Monogr. Phys.} {\bfseries 103} (1999) 1}.

\bibitem{Branco:2011iw}
G.~C. Branco, P.~M. Ferreira, L.~Lavoura, M.~N. Rebelo, M.~Sher and J.~P.
  Silva, \emph{{Theory and phenomenology of two-Higgs-doublet models}},
  \href{https://doi.org/10.1016/j.physrep.2012.02.002}{\emph{Phys. Rept.}
  {\bfseries 516} (2012) 1} [\href{https://arxiv.org/abs/1106.0034}{{\ttfamily
  1106.0034}}].

\bibitem{Ivanov:2017dad}
I.~P. Ivanov, \emph{{Building and testing models with extended Higgs sectors}},
  \href{https://doi.org/10.1016/j.ppnp.2017.03.001}{\emph{Prog. Part. Nucl.
  Phys.} {\bfseries 95} (2017) 160}
  [\href{https://arxiv.org/abs/1702.03776}{{\ttfamily 1702.03776}}].

\bibitem{Freitas:2020ttd}
F.~F. Freitas, J.~Gonçalves, A.~P. Morais and R.~Pasechnik,
  \emph{{Phenomenology of vector-like leptons with Deep Learning at the Large
  Hadron Collider}},  \href{https://arxiv.org/abs/2010.01307}{{\ttfamily
  2010.01307}}.

\end{thebibliography}\endgroup

\end{document}